\documentclass[a4paper,fleqn,usenatbib]{mnras}

\usepackage{newtxtext,newtxmath}
\usepackage[T1]{fontenc}
\usepackage{ae,aecompl}

\usepackage{graphicx}	% Including figure files
\usepackage{amsmath}	% Advanced maths commands
\usepackage{amssymb}	% Extra maths symbols
\usepackage{pdflscape}
\usepackage{array}

\usepackage{xspace}
\newcommand{\geu}{iPTF16geu\xspace}
\newcommand{\scipap}{G17\xspace}
\newcommand{\sn}{SN\xspace}
\newcommand{\snia}{SN~Ia\xspace}

\newcommand{\glsne}{glSNe\xspace}

\newcommand{\wfc}{WFC3\xspace}

\newcommand{\wfcuvis}{WFC3/UVIS\xspace}
\newcommand{\wfcir}{WFC3/IR\xspace}
\newcommand{\uvisaperture}{{\tt UVIS2-C512C-SUB}\xspace}
\newcommand{\iraperture}{{\tt IRSUB512}\xspace}
\newcommand{\jband}{{\it J}\xspace}
\newcommand{\hband}{{\it H}\xspace}
\newcommand{\ksband}{{\it Ks}\xspace}
\newcommand{\hstu}{$F390W$\xspace}
\newcommand{\hstb}{$F475W$\xspace}
\newcommand{\hstr}{$F625W$\xspace}
\newcommand{\hsti}{$F814W$\xspace}
\newcommand{\hstj}{$F110W$\xspace}
\newcommand{\hsth}{$F160W$\xspace}

\newcolumntype{H}{>{\setbox0=\hbox\bgroup}c<{\egroup}@{}}

\title[Magnification, dust and time-delay from \geu]{Magnification, dust and time-delay constraints from the first resolved strongly lensed Type Ia supernova \geu}

\author[Dhawan et al.]{%
S.~Dhawan,$^{1}$ \thanks{E-mail: suhail.dhawan@fysik.su.se}
J.~Johansson,$^{2}$
A.~Goobar,$^{1}$
R.~Amanullah,$^{1}$
E.~M{\"o}rtsell$^{1}$
\newauthor
S.B.~Cenko,$^{3,4}$
A.~Cooray,$^{5}$
O.~Fox,$^{6}$
D. Goldstein,$^{7\thanks{Hubble Fellow.}}$
R.~Kalender,$^{1}$
M.~Kasliwal,$^{7}$
\newauthor
S.R. Kulkarni,$^{7}$
W.H. Lee,$^{8}$
H.~Nayyeri,$^{5}$
P. Nugent,$^{9,10}$
E. Ofek,$^{11}$
R. Quimby$^{12,13}$
\\
$^{1}$The Oskar Klein Centre, Physics Department,
    Stockholm University,
    Albanova University Center, SE 106 91 Stockholm, Sweden\\
$^{2}$Department of Physics and Astronomy, Division of Astronomy and Space Physics, Uppsala University, Box 516, SE 751 20 Uppsala, Sweden\\
$^{3}$Astrophysics Science Division, NASA Goddard Space Flight Center, 8800 Greenbelt Road, Greenbelt, MD 20771, USA\\
$^{4}$Department of Astronomy, University of Maryland, College Park, MD 20742, USA\\
$^{5}$Department of Physics \& Astronomy, University of California, Irvine, CA 92697, USA\\
$^{6}$Space Telescope Science Institute, 3700 San Martin Dr., Baltimore, MD 21218, USA\\
$^{7}$Division of Physics, Mathematics and Astronomy, California Institute of Technology, Pasadena, CA 91125, USA\\
$^{8}$Instituto de Astronom{\'i}a, Universidad Nacional Aut{\'o}noma de M{\'e}xico, Apdo. Postal 70-264, Cd. Universitaria, Cd de M{\'e}xico, 04510, Mexico \\
$^{9}$Department of Astronomy, University of California, Berkeley, CA 94720-3411, USA\\
$^{10}$Lawrence Berkeley National Laboratory, 1 Cyclotron Road, MS 50B-4206, Berkeley, CA
94720, USA\\
$^{11}$Benoziyo Center for Astrophysics, Weizmann Institute of Science, 76100 Rehovot, Israel\\
$^{12}$ Department of Astronomy/Mount Laguna Observatory, San Diego State University, 5500 Campanile Drive, San Diego, CA 92812-1221, USA\\
$^{13}$ Kavli IPMU (WPI), UTIAS, The University of Tokyo, Kashiwa, Chiba 277-8583, Japan \\  
}

\date{Accepted XXX. Received YYY; in original form ZZZ}

\pubyear{2019}

\begin{document}
\label{firstpage}
\pagerange{\pageref{firstpage}--\pageref{lastpage}}
\maketitle

% Abstract of the paper
\begin{abstract}
We report lensing magnifications,  extinction, and time-delay estimates for the first resolved, multiply-imaged Type Ia supernova iPTF16geu, at $z = 0.409$, using \emph{Hubble Space Telescope} ($HST$) observations in combination with supporting ground-based data. Multi-band photometry of the resolved images provides unique information about the differential dimming due to dust in the lensing galaxy. Using $HST$ and Keck AO reference images taken after the SN faded, we obtain a total lensing magnification for iPTF16geu of  $\mu = 67.8^{+2.6}_{-2.9}$, accounting for extinction in the host and lensing galaxy.
As expected from the symmetry of the system, we measure very short  time-delays for the three fainter images with respect to the brightest one: -0.23 $\pm$ 0.99, -1.43 $\pm$ 0.74 and 1.36 $\pm$ 1.07 days.  Interestingly, we find large differences between the magnifications of the four supernova images, even after accounting for uncertainties in the extinction corrections:  $\Delta m_1 = -3.88^{+0.07}_{-0.06}$,   $\Delta m_2 =  -2.99^{+0.09}_{-0.08}$,  $\Delta m_3 = -2.19^{+0.14}_{-0.15}$ and  $\Delta m_4 = -2.40^{+0.14}_{-0.12}$ mag,  discrepant with model predictions suggesting similar image brightnesses. A possible explanation for the large differences is gravitational lensing by substructures, micro- or millilensing, in addition to the large scale lens causing the image separations. We find that the inferred magnification is insensitive to the assumptions about the dust properties in the host and lens galaxy. 
\end{abstract}
%hej hej
% Select between one and six entries from the list of approved keywords.
% Don't make up new ones.

%%%%%%%%%%%%%%%%%%%%%%%%%%%%%%%%%%%%%%%%%%%%%%%%%%

%%%%%%%%%%%%%%%%% BODY OF PAPER%%%%%%%%%%%%%%%%%%

\section{Introduction}
The discovery of the first multiply-imaged gravitationally lensed Type Ia supernova (SN~Ia), iPTF16geu \citep[][hereafter G17]{2017Sci...356..291G} was a major breakthrough for time-domain astronomy, highlighting the power of wide-field surveys to detect rare phenomena.  Transient astrophysical sources that are strongly lensed by foreground galaxies or galaxy clusters are powerful probes in cosmology since they make it possible to measure time delays between the multiple images. More than half a century has passed since \citet{1964MNRAS.128..307R} proposed that time-delays between multiple images of transients like supernovae (SNe) are useful since they depend sensitively on cosmological parameters, e.g. the Hubble constant ($H_0$). The observations also allow us to probe the distribution of matter in the lens. Hence, multiply-resolved gravitationally lensed supernovae (\glsne) are exquisite laboratories for fundamental physics, as well as astrophysical properties of the host and lens galaxies  \citep[see][for a review of strongly lensing of SNe and other explosive transients]{2019arXiv190706830O}. Although strongly lensed galaxies and quasars are more common than \glsne, \glsne have notable advantages, particularly if they are of Type Ia (glSNe~Ia). This is because the ``standard candle" nature of the SNe~Ia allows us to directly measure the magnification factor, which helps us to overcome various degeneracies in estimating $H_0$ from strongly lensed transients, including the mass-sheet degeneracy \citep{1985ApJ...289L...1F,2014A&A...564A.103S}. Additionally, SNe~Ia have a well-studied family of light curves and hence, can be used for an accurate measurement of time-delays, with significantly fewer follow-up observations than quasars. Moreover, since SNe fade away, we can obtain post-explosion imaging to validate the lens model.

However, discovering these rare events has proven very challenging and it is only thanks to the recent developments in time domain astronomy that the first lensed supernovae have been detected. \citet{2013ApJ...768L..20Q} found a strongly lensed SN~Ia but multiple images were not resolved. However, the magnification allowed for high signal to noise spectroscopy at $z > 1$ and hence, the SN was used to show that the spectral properties of high-$z$ and nearby SNe~Ia are very similar \citep{2017A&A...603A.136P}.  
The first resolved lensed supernova, SN Refsdal \citep{2015Sci...347.1123K}, detected with the Hubble Space Telescope ($HST$), is a core-collapse \sn magnified by a cluster of galaxies, which makes the lens modelling challenging \citep{2018ApJ...860...94G}. 

The discovery of iPTF16geu showed that glSNe~Ia can be found without the need of highly spatially resolved observations, thanks to their "standard candle" nature. At $z=0.409$, the \sn was found to be 30 standard deviations too bright compared with the \snia population, which prompted us to observe the system from space with $HST$ and with laser guide star adaptive optics (LGS-AO) at VLT and Keck. Here we report on the multi-wavelength follow-up observations carried out while the \sn was active in late 2016, as well as laser aided AO Near IR observations with Keck and $HST$ observations after the SN had faded below the detection limit.
We use the multi-wavelength light curves from the resolved images in combination with unresolved, ground-based data to constrain the magnifications of the individual SN images (and hence, the total SN magnification) after accounting for the extinction in the different lines of sight to the multiple images.

The accurate \sn image positions are used to model the lens, as described in an accompanying paper (M{\"o}rtsell et al, in prep). Unlike the case of strongly lensed quasars, as the transient faded, we had an opportunity to verify the lensing model with the reconstruction of the distorted host galaxy image. 
We compared the model predictions of the flux ratios (based only on the image positions) 
to the observed value after extinction correction for each image. We used this to assess the possibility that otherwise unaccounted for residuals are caused due to lensing by substructures within the lensing galaxy, in the form of field stars.  
In another accompanying paper (Johansson et al, in prep) we present the spectroscopic observations of \geu.

The structure of this paper is as follows. We present the observations of iPTF16geu in Section~\ref{sec:data} and the photometric analysis method in Section~\ref{sec:photometry}. In Section~\ref{sec:lcmodel}, we describe the multiple-image SN model and the resulting lensing magnification, time-delays and constraints on extinction properties in Section~\ref{sec:results}. We present the observed magnifications in context of model predictions and discuss the possibility of substructures in Section~\ref{sec:substructure} and discuss observations of future strongly lensed SNe in Section~\ref{sec:forecast}. Finally, we present our conclusions in Section~\ref{sec:conclusion}.
\section{Observations}
\label{sec:data}
\subsection{Hubble Space Telescope Wide Field Camera 3}
We observed \geu with the Hubble Space Telescope ($HST$) Wide Field Camera~3 (\wfc) under programs DD~14862 and 
GO~15276 (PI: A.~Goobar) using the ultra violet (\wfcuvis) and near-IR (\wfcir) channels.  For both channels we only read out part, 
$512\times512$ pixels, of the full detectors.  However, given the different pixel scales of the two channels, 
$0.04''$/pixel and $0.12''$/pixel for \wfcuvis and \wfcir, respectively, they will not cover the same area on the sky.  
The data were obtained using either a 3- or 4-point standard dithering pattern for both channels.  For \wfcuvis we used 
the \uvisaperture, that is located next to the amplifier, with a post-flash to maximize the charge transfer efficiency (CTE) 
during read-out.  All imaging data are shown in Table~\ref{tb:hstimaging}.

%% Table of HST imaging data
\begin{table}
\centering
\caption{Hubble Space Telescope Wide Field Camera 3 data imaging data presented here.  The columns are the civil date, 
the Modified Julian Date (MJD), the $HST$ passband, total exposure time, the number of sub-exposures, and the \wfc camera.  
The \wfcuvis and \wfcir data were obtained with the  \uvisaperture and \iraperture subarray, respectively.  \label{tb:hstimaging}}
\begin{tabular}{cccccc}
\hline\hline
Civil date & MJD & Filter & Exp. & Sub & Camera \\
\hline
2016-10-20 & 57681.62 & $F475W$ & 378.0 & 3 & UVIS2 \\
2016-10-20 & 57681.62 & $F625W$ & 291.0 & 3 & UVIS2 \\
2016-10-20 & 57681.63 & $F814W$ & 312.0 & 3 & UVIS2 \\
2016-10-20 & 57681.64 & $F110W$ & 63.9 & 3 & IR \\
2016-10-20 & 57681.64 & $F160W$ & 621.4 & 3 & IR \\
2016-10-25 & 57685.91 & $F625W$ & 198.0 & 3 & UVIS2 \\
2016-10-25 & 57685.92 & $F814W$ & 114.0 & 3 & UVIS2 \\
2016-10-25 & 57685.93 & $F475W$ & 183.0 & 3 & UVIS2 \\
2016-10-25 & 57685.94 & $F390W$ & 429.0 & 3 & UVIS2 \\
2016-10-25 & 57685.99 & $F110W$ & 63.9 & 3 & IR \\
2016-10-25 & 57685.99 & $F160W$ & 415.1 & 3 & IR \\
2016-10-29 & 57689.89 & $F475W$ & 378.0 & 3 & UVIS2 \\
2016-10-29 & 57689.91 & $F625W$ & 291.0 & 3 & UVIS2 \\
2016-10-29 & 57689.91 & $F814W$ & 312.0 & 3 & UVIS2 \\
2016-10-29 & 57689.96 & $F110W$ & 63.9 & 3 & IR \\
2016-10-29 & 57689.96 & $F160W$ & 621.4 & 3 & IR \\
2016-11-02 & 57694.21 & $F625W$ & 804.0 & 4 & UVIS2 \\
2016-11-02 & 57694.21 & $F814W$ & 480.0 & 4 & UVIS2 \\
2016-11-02 & 57694.24 & $F110W$ & 63.9 & 3 & IR \\
2016-11-02 & 57694.24 & $F160W$ & 415.1 & 3 & IR \\
2016-11-02 & 57694.28 & $F105W$ & 309.4 & 3 & IR \\
2016-11-06 & 57698.25 & $F625W$ & 644.0 & 4 & UVIS2 \\
2016-11-06 & 57698.25 & $F814W$ & 420.0 & 4 & UVIS2 \\
2016-11-06 & 57698.26 & $F110W$ & 63.9 & 3 & IR \\
2016-11-06 & 57698.26 & $F160W$ & 621.4 & 3 & IR \\
2016-11-10 & 57702.16 & $F625W$ & 804.0 & 4 & UVIS2 \\
2016-11-10 & 57702.16 & $F814W$ & 480.0 & 4 & UVIS2 \\
2016-11-10 & 57702.17 & $F110W$ & 63.9 & 3 & IR \\
2016-11-10 & 57702.17 & $F160W$ & 415.1 & 3 & IR \\
2016-11-15 & 57707.12 & $F625W$ & 644.0 & 4 & UVIS2 \\
2016-11-15 & 57707.12 & $F814W$ & 420.0 & 4 & UVIS2 \\
2016-11-15 & 57707.14 & $F110W$ & 63.9 & 3 & IR \\
2016-11-15 & 57707.14 & $F160W$ & 621.4 & 3 & IR \\
2016-11-17 & 57709.72 & $F625W$ & 804.0 & 4 & UVIS2 \\
2016-11-17 & 57709.74 & $F814W$ & 480.0 & 4 & UVIS2 \\
2016-11-17 & 57709.78 & $F110W$ & 63.9 & 3 & IR \\
2016-11-17 & 57709.78 & $F160W$ & 415.1 & 3 & IR \\
2016-11-22 & 57714.32 & $F625W$ & 644.0 & 4 & UVIS2 \\
2016-11-22 & 57714.32 & $F814W$ & 420.0 & 4 & UVIS2 \\
2016-11-22 & 57714.35 & $F110W$ & 63.9 & 3 & IR \\
2016-11-22 & 57714.35 & $F160W$ & 621.4 & 3 & IR \\
2018-11-10 & 58432.37 & $F390W$ & 1454.0 & 4 & UVIS2 \\
2018-11-10 & 58432.38 & $F475W$ & 1494.0 & 4 & UVIS2 \\
2018-11-10 & 58432.40 & $F814W$ & 1227.0 & 4 & UVIS2 \\
2018-11-10 & 58432.40 & $F625W$ & 1648.0 & 4 & UVIS2 \\
2018-11-10 & 58432.48 & $F110W$ & 125.3 & 3 & IR \\
2018-11-10 & 58432.48 & $F105W$ & 483.9 & 3 & IR \\
2018-11-10 & 58432.48 & $F160W$ & 965.3 & 3 & IR \\
\hline\hline
\end{tabular}

\end{table}

We used the automatic \texttt{calwf3} reduction pipeline at the Space Telescope Science Institute, on all the data to dark subtract, flat-field and correct the data for charge-transfer inefficiency.  The individual images were then combined and corrected for geometric distortion using the \texttt{AstroDrizzle} software\footnote{https://wfc3tools.readthedocs.io}.

\subsection{Ground data}
In addition to the data already presented in \scipap, \geu was observed from the ground until it disappeared behind the Sun. We obtained laser guided adaptive optics (LGS-AO) observations with NIRC2 at the Keck~II telescope on Mauna Kea in \jband, \hband and $K_S$ bands on UTC 2017, June, 16, after \geu had faded.  For the 
\jband and \hband bands we obtained 9~exposures in a dithering pattern, each with an integration time of 20\,s.  For the $K_S$ band, 
18~exposures of 65\,s were acquired. 

Standard near infrared reduction was applied where the individual images were first dark subtracted and flat fielded.  The flat 
frames were obtained using the same dome on-off technique as described in \scipap.  The sky background for each science 
frame was obtained from the images preceding and following each exposure, after first masking out the object.  
Together with data presented in \scipap this resulted in a total of 3~epochs for the NIRC2/\jband and 2~epochs for the 
NIRC2/\hband and NIRC2/\ksband bands, respectively.

We also obtained $rizYJH$ photometry of iPTF16geu with the multi-channel Reionization And Transients InfraRed camera \citep[RATIR;][]{butler2012} mounted on the 1.5-m Johnson telescope at the Mexican Observatorio Astronomico Nacional on Sierra San Pedro Martir (SPM) in Baja California, Mexico \citep{watson2012}.  
The RATIR data were reduced and coadded using standard CCD and IR processing techniques in IDL and Python, utilizing the online astrometry programs SExtractor and SWarp. Calibration was performed using field stars with reported fluxes in both 2MASS \citep{2006AJ....131.1163S} and the SDSS Data Release 9 Catalogue \citep{2012ApJS..203...21A}. % Template image subtracted ...
%%% PHOTOMETRY
%%%
\section{Photometric analysis}
\label{sec:photometry}
In this section we detail the analysis methodology to obtain multiband WFC3 photometry for the resolved images. In section~\ref{sec:nirc2} we detail the procedure to forward model the NIRC2 LGS-AO images. We derive the WFC3 photometry using two different approaches. The forward modelling approach is described in section~\ref{sec:wfc3_forwardmodel} and the template subtractions in section~\ref{ssec:phot_sub_joel}. We measure the fluxes for all the SN images simultaneously. For our analyses, we use the template subtracted photometry since this approach is independent of the assumptions on the host and lens galaxy models.

\subsection{Forward modelling of the NIRC2 images}\label{sec:nirc2}
The LGS-AO NIRC2 images have the highest spatial resolution in our data set.  We use the NIRC2 data to build a 
parametric model of the \geu system, including the \sn images, the host galaxy, and the lens. The model we use is described in detail in \S\ref{sec:bgmodel}, and is only briefly summarized here. The shape of the lensing galaxy is modeled 
with a S\'ersic profile \citep{1963BAAA....6...41S} while the \sn images are 
modeled by the point-spread function (PSF) of the images, assumed to be a Moffat profile.  The shape of the \sn host galaxy is described by the expansion 
in eq.~\eqref{eq:host}.  The full model is then fitted simultaneously to all data in one NIRC2 filter 
at a time.  Some parameters,  such as the host and lens models and the position of the \sn images, are forced to be 
the same for all available images in a filter, while the fluxes of the \sn images are allowed to vary between the 
different epochs, with the exception of the reference images obtained in 2017 where all \sn fluxes are fixed to zero, which breaks the
degeneracy between the PSFs and the background model.

\begin{figure*}
	\centering
	\includegraphics[width=0.95\textwidth]{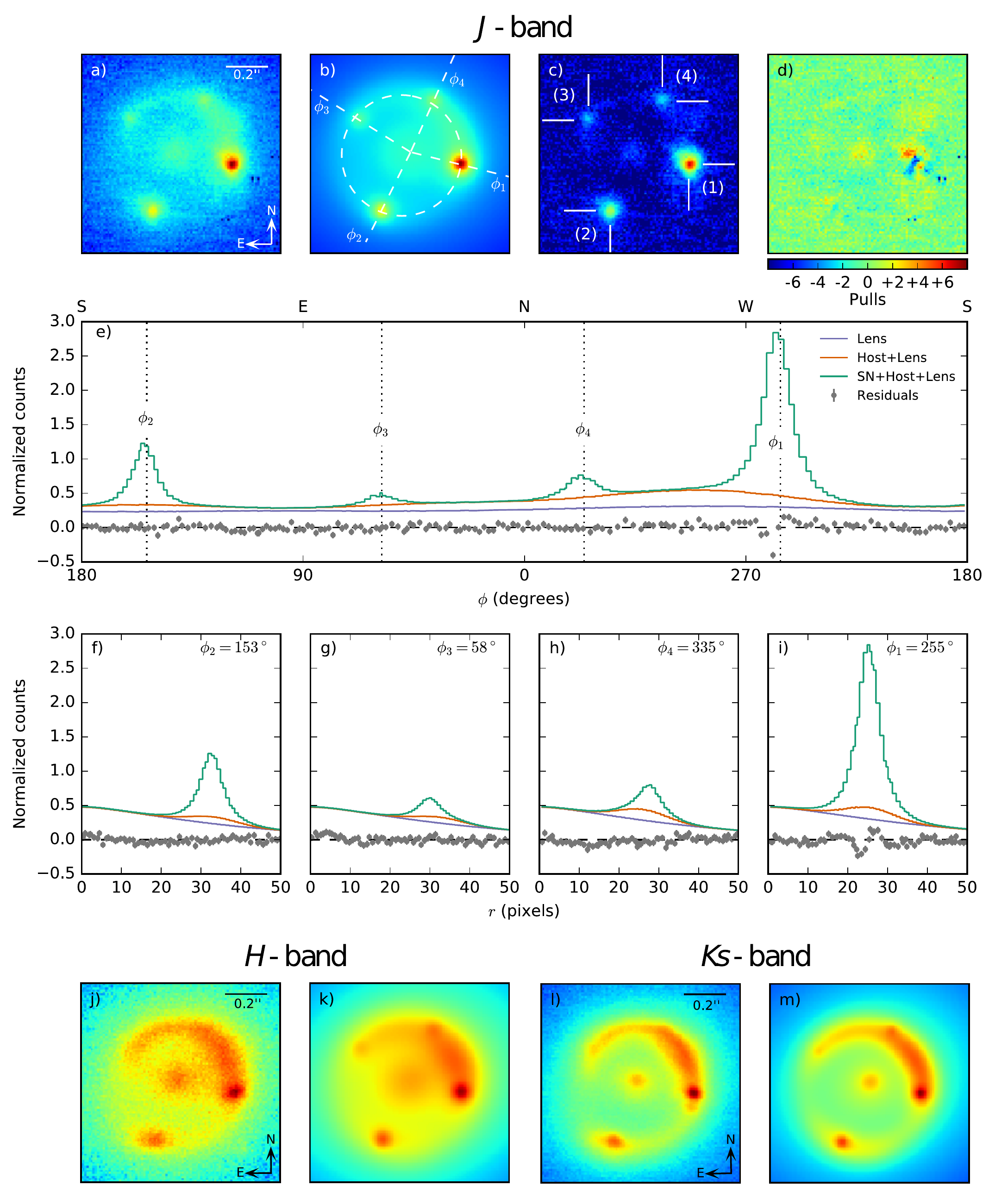}
	\caption{%
		{\bf a)} NIRC2 \jband image of the \geu system obtained on on Nov 5, 2016. 
		{\bf b)} The model fitted simultaneously to all available epochs as described in the text.  The dashed circle shows the 
		position of the host galaxy as described in eq.~\eqref{eq:hostradius}.  The dashed lines are showing the angular 
		positions of the four \sn images. 
		{\bf c)} The subtraction between the data and the host and lens models.  The fitted PSF positions of the four \sn images 
		have been marked.  
		{\bf d)} The ''pulls'', i.e. the residuals normalized with the pixel uncertainties when the lens, host and \sn 
		model is subtracted from the data.  
		{\bf e)} The profile of both the model and the residuals along the host radius marked by the dashed circle in b), The 
		fitted angles, $\phi_i$, of the \sn images are marked by the dotted, black lines.
		{\bf f)} -- {\bf i)}  The radial profiles from the center for the \sn images as marked and labelled in b).
		{\bf j)} -- {\bf k)} NIRC2 \hband image obtained on Oct 23, 2016 and the corresponding fitted model.
		{\bf l)} -- {\bf m)} NIRC2 \ksband image obtained on Oct 22, 2016 and the corresponding fitted model.
	\label{fig:nirc2}}
\end{figure*}

Examples of data and the fitted models are shown in Figure~\ref{fig:nirc2}.  The fitted positions of the four \sn
images for the NIRC2 \jband-band are further presented Table~\ref{tb:snpos}, while the lens and host parameters are shown 
in Tables~\ref{tb:lensmodel}--\ref{tb:hostmodelwidth}. We use the \jband-band as the reference since the ratio between the \sn flux and background is the highest of all NIRC2 filters and we have two epochs where the \sn is active.
% The fitted \sn fluxes are presented in Table~\ref{tb:snflux}.  The PSFs  were placed at the locations along the fitted host 
% galaxy model ($r_H(\phi)$ in Eq.~\eqref{eq:host}), while avoiding the \sn images.  We do not attempt to calibrate the fluxes 
% due to the absence of reference objects, in particular stars, in the NIRC2 images.

%% PLOT: NIRC2 fit with residuals and profile plots

%% TABLE: Host widths
\begin{table}
	\centering
	\caption{%
		The fitted widths, $\sigma_H$, for the host model defined in eq.~\eqref{eq:host}.
	\label{tb:hostmodelwidth}}
	\begin{tabular}{l | r @{\hspace{0.5em}} l}
\hline\hline
Filter & \multicolumn{2}{c}{$\sigma_H$}\\
  & \multicolumn{2}{c}{$('')$}\\
\hline
$K_s$ & 0.080 & (0.001)\\
$H$ & 0.129 & (0.001)\\
$F160W$ & $^*$0.129 & \\
$J$ & 0.102 & (0.003)\\
$F110W$ & $^*$0.102 & \\
$F814W$ & 0.067 & (0.002)\\
$F625W$ & 0.056 & (0.003)\\
$F475W$ & $^*$0.056 & \\
\hline\hline
\end{tabular}
	
\end{table}

As seen in Figure~\ref{fig:nirc2}, the model generally fits the data well.  Four \sn images are clearly visible in the \jband-band 
but from panel~d) we also see that the fit is not perfect.  Discrepancies can mainly be seen for the brightest \sn image, which are 
probably due to an imperfect PSF model, rather than an insufficient background model.  In other words, if the systematic PSF
uncertainties are known, the method can be used to obtain fluxes for the four \sn images.

From the radial profile plots in panels~f)--i) there is an apparent degeneracy between the host model and the \sn profiles given 
that their extrema coincide and have similar width.  However, recall that we only fit one parameter, $\sigma_H$, for the width of 
the host model, as explained in \S\ref{sec:bgmodel}, and the value of this parameter will mainly be determined by the pixels 
between the images located along the dashed circle in panel~b).  Furthermore, by studying the profile along this circle, as shown in 
panel~e), we can conclude that the maximum of the host galaxy amplitude appears to be located between images (1) and (4),  
and the best fit model suggest that the background flux under the \sn images is either increasing or decreasing monotonically.

Since the image positions are not expected to change with time or wavelength, we fixed the positions to the 
values in Table~\ref{tb:snpos} for the remaining of the analysis in this paper.  %Using the \jband as the reference, is motivated 
%both by the fact that the ratio between the \sn flux and the background is higher than for the other NIRC2 filters, and that we 
%have two epochs where the \sn is active.

With the \sn positions fixed we move on to fit the host model for the remaining NIRC2 filters.  The fitted models to the 
\hband- and \ksband-band are shown in panels j)--m) in Figure~\ref{fig:nirc2}.  In the figure, the epochs when the \sn was 
active are shown together with the corresponding data.  The lens and host model parameters are presented in 
Tables~\ref{tb:lensmodel}--\ref{tb:hostmodelwidth}.  

 Note that in this paper, we number the Images 1 $\rightarrow$ 4 clockwise from the brightest image (middle right in the data presented in  Figure~\ref{fig:nirc2} Panel c))
% Similar to the \sn positions we do not expect the position of the host galaxy model, $r_H(\phi)$, to change significantly 
% between different filters.

%% Why is there a difference between the Sersic index for the UVIS and NIRC2?  Is it expected to be wavelength dependent?
%% Close to an exponential profile nS = 1, "which is a good description of spiral galaxy disks and dwarf elliptical galaxies."
\iffalse
\begin{figure*}
	\centering

	\includegraphics[width=\textwidth]{wfc3_patches_v2.pdf}
	\caption{%
		Examples of data, model and residual patches for each \wfc filter.  The residual patches have been normalised by the 
		pixel uncertainty to form the "pull".  For each filter, the epoch where the \sn was the brightest is shown.  This corresponds
		to Oct.~25, 2016 for \hstu, \hstb, \hstr, and \hsti,  and Nov.~2nd, 2016 for \hstj and \hsth, respectively. In this paper, we number the Images 1$\rightarrow$4 clockwise from the brightest Image in the top right corner of the observations. 
	\label{fig:wfcforward}}
\end{figure*}
\fi
\begin{figure*}
    \centering
    \includegraphics[width=\textwidth,trim={0 10mm 0 8mm},clip]{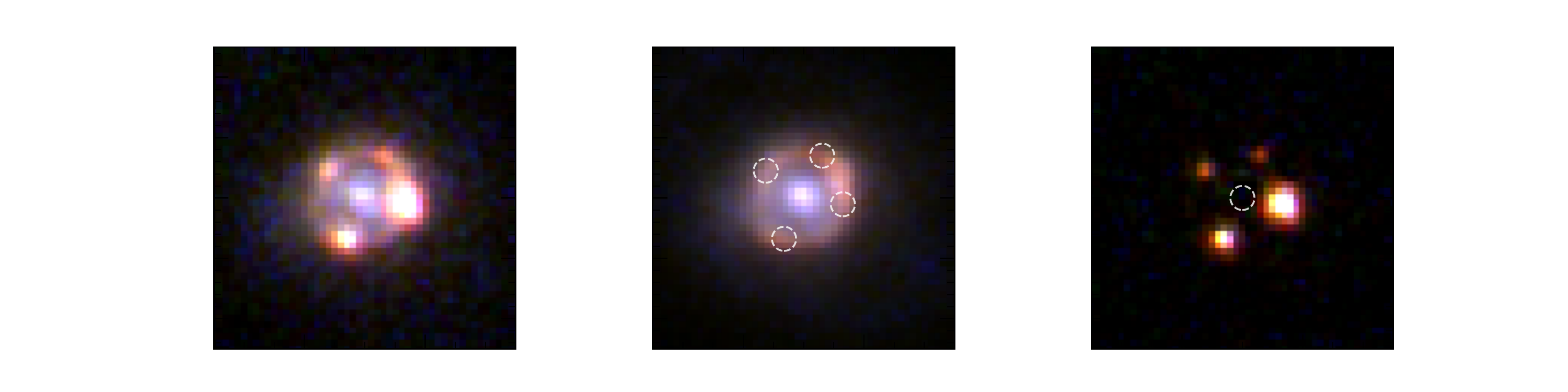}
    \caption{$HST$ observations from 2016, October, 29, in a combination of $F814W$, $F625W$, $F475W$ filters of iPTF16geu (left), post-explosion template (middle) and the subtracted image (right) }
    \label{fig:subtractions_rgb}
\end{figure*}

\subsection{WFC3 photometry using forward modelling}
\label{sec:wfc3_forwardmodel}
Here we describe the WFC3 photometry computed using the forward modelling approach.  We emphasize that this procedure was only used to extract fluxes \emph{before} the images after the SN faded were obtained. The photometry estimated using this method is not used in any of the analyses described below.
We use the same approach detailed in \S\ref{sec:nirc2} to extract the SN photometry.  Details of the fitting procedure are described in \S~\ref{sec:wfcpsf} and the fitted parameters presented in Tables~\ref{tb:lensmodel}--\ref{tb:hostmodelwidth}.  
%We use the approach detailed in \S\ref{sec:nirc2} to extract the  \sn photometry. However, given the lower resolution and the varying signal-to-noise of the host and lensing galaxy, all parameters could not be fitted for every filter. Similar to the NIRC2 data, the field of view for \wfcuvis is too small and the data do not include any isolated stars that can be used to model the PSF.
%While conditions in space are relatively stable, the PSF of \wfc varies over both the focal 
%plane and in time.  However, here we use a simple approach of simultaneously fitting the parameters of a Moffat profile (see Section~\ref{sec:wfcpsf} for details). For \wfcuvis we fit the parameters 
%to bright stars observed over time with the same subarray as \geu. Since the field-of-view for the \wfcir data is larger, and the bright star $20\arcsec$ North of \geu, was included in all observations.

  Below we describe the procedure to build lightcurves from template subtracted images. Since we see that there are some significant residuals in panel d) of Figure~\ref{fig:nirc2} and that the template subtraction approach is significantly more model independent, we use the resulting SN fluxes from the template subtractions in our analyses.

\begin{figure*}
\centering
\includegraphics[width=\textwidth,trim={0 10mm 0 8mm},clip]{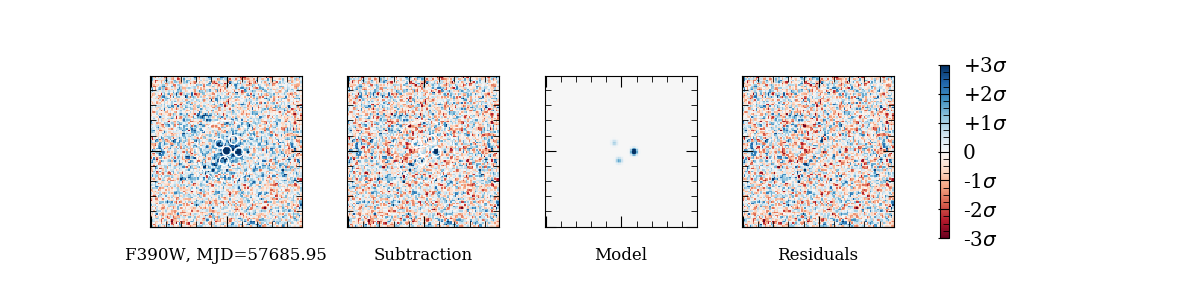}
\includegraphics[width=\textwidth,trim={0 10mm 0 8mm},clip]{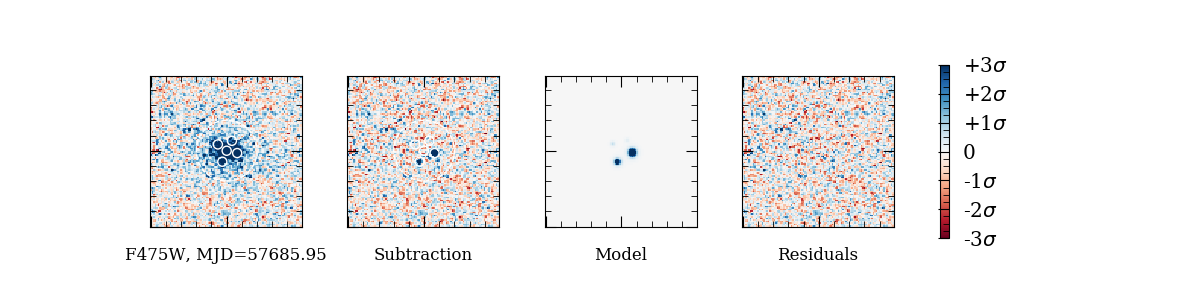}
\includegraphics[width=\textwidth,trim={0 10mm 0 8mm},clip]{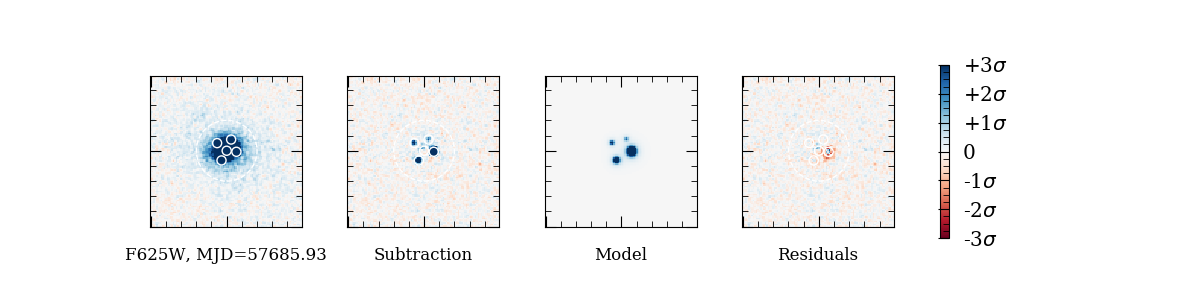}
\includegraphics[width=\textwidth,trim={0 10mm 0 8mm},clip]{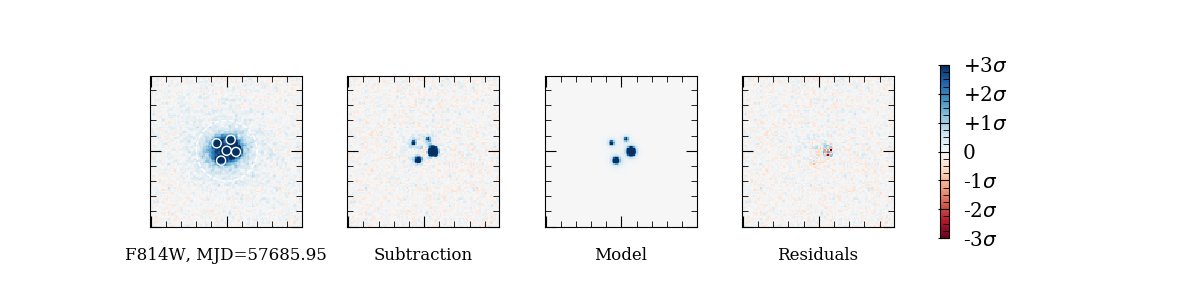}
\includegraphics[width=\textwidth,trim={0 10mm 0 8mm},clip]{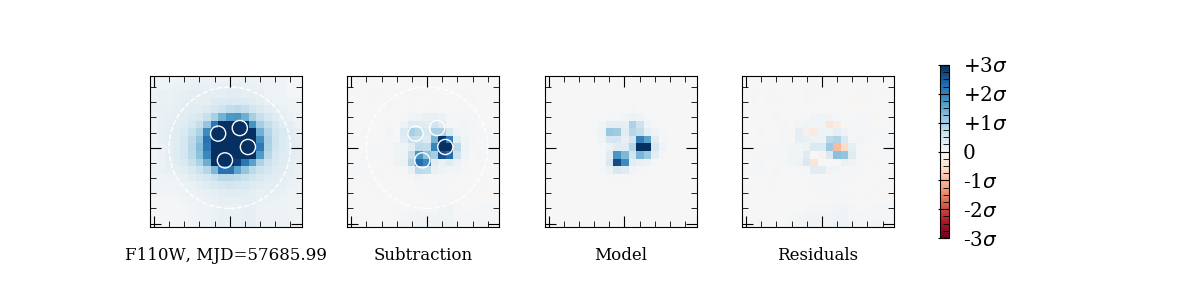}
\includegraphics[width=\textwidth,trim={0 10mm 0 8mm},clip]{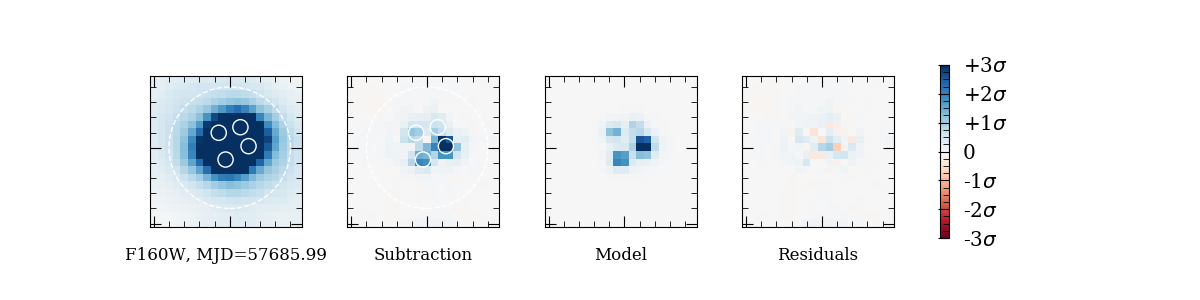}
\caption{
	$HST$ observations taken on 2016 October 20 in the $F390W$, $F475W$, $F625W$, $F814W$, $F110W$, $F160W$ filters, together with template subtracted images, fitted PSF models and residuals after the four SN PSFs have been subtracted, for each filter.
	\label{fig:subtractions}
}
\end{figure*}

 \begin{table*}
\centering
\caption{This table summarises the resulting best fit parameters by fitting an \texttt{sncosmo} model to the ground-based and HST observations (equation~\ref{eq:chi}). For each image, we present the time of maximum, t$_{max}$ (relative to image 1), host galaxy colour excess (which is treated as being the same for each image), the lens galaxy colour excess as well as the magnification for each image (given as the 68$\%$ credible region of the posterior distribution). Our fiducial case assumes $R_V =2$ in both the host and the lens galaxy. We also present the parameters values assuming host $R_V =2$ and lens $R_V$ as a free parameters as well as the case with host and lens $R_V$ fixed to the Milky Way value of 3.1.}
\resizebox{\textwidth}{!}{\begin{tabular}{r | c c c c}
\hline
Parameter & Image 1 & Image 2 &  Image 3 & Image 4 \\
\hline\hline
{\bf Fiducial fit parameters:} & & & & \\
(fixed $R_{V}^{\rm lens} \equiv 2$, fixed $R_{V}^{\rm host} \equiv 2$) & & & &  \\
$t_{\rm max}$ & {\bf 57652.80 ($\pm$ 0.33)} & { -0.23 ($\pm$  0.99)} & { -1.43 ($\pm$ 0.74)} & { 1.36 ($\pm$ 1.07)} \\
{\bf Stretch, $s$} & {\bf 0.99 ($\pm$ 0.01)} & {\bf Same as Image 1}  & {\bf Same as Image 1}  & {\bf Same as Image 1}  \\
{\bf $E(B-V)_{\rm host}$}   &  {\bf 0.29 ($\pm$ 0.05)} &   {\bf Same as Image 1} & {\bf Same as Image 1} & {\bf Same as Image 1}  \\
{\bf $E(B-V)_{\rm lens}$}   &  {\bf 0.06 ($\pm$ 0.08)} &  {\bf 0.17 ($\pm$ 0.08)} & {\bf 0.42 ($\pm$ 0.09)} & {\bf 0.94 ($\pm$ 0.07)} \\
{\bf Magnification} & {\bf $-3.88^{+0.07}_{-0.06}$} & {\bf $-2.99^{+0.09}_{-0.08}$} & {\bf $-2.19^{+0.14}_{-0.15}$} & {\bf $-2.40^{+0.14}_{-0.12}$ }
\\ \hline
{\bf Reddening assumptions modified:} & & & & \\
(free $R_{V}^{\rm lens}$, fixed $R_{V}^{\rm host} \equiv 2$) & & & & \\
$t_{\rm max}$  &  57652.9 ($\pm$ 0.20) & -0.31 ($\pm$  0.93) & -1.84  ($\pm$ 0.90) & 0.77 ($\pm$ 1.27) \\
Stretch, $s$ & 1.00 ($\pm$ 0.01) & Same as Image 1 & Same as Image 1 & Same as Image 1  \\
$E(B-V)_{\rm host}$  & 0.18 ($\pm$ 0.05) & Same as Image 1 & Same as Image 1 & Same as Image 1  \\
$E(B-V)_{\rm lens}$  &  0.26 ($\pm$ 0.09) &  0.41 ($\pm$ 0.10) & 0.78 ($\pm$ 0.12) & 1.58 ($\pm$ 0.14) \\
$R_{V}^{\rm lens}$ (single $R_V$) & $< 1.8 $ & Same as Image 1 & Same as Image 1 & Same as Image 1 \\
$R_{V}^{\rm lens}$ (all free) & $< 2.8 $ & $< 3.2 $ & $< 3.6$ & $< 1.5 $ \\
Magnification & $-3.78^{+0.09}_{-0.10}$ & $-2.85^{+0.10}_{-0.11}$ & $-1.86^{+0.15}_{-0.15}$ & $-1.98^{+0.16}_{-0.17}$ \\
\hline
($R_{V}^{\rm lens} \equiv 3.1$,  $R_{V}^{\rm host} \equiv 3.1$) \\
$t_{\rm max}$  &  57652.7 ($\pm$ 0.38) & 0.11 ($\pm$  0.91) & -0.85  ($\pm$ 1.08) & -0.41 ($\pm$ 2.27) \\
Stretch, $s$ & 1.01 ($\pm$ 0.01) & Same as Image 1 & Same as Image 1 & Same as Image 1 \\
$E(B-V)_{\rm host}$  & 0.17 ($\pm$ 0.08) & Same as Image 1 & Same as Image 1 & Same as Image 1\\
$E(B-V)_{\rm lens}$  &  0.13 ($\pm$ 0.08) &  0.20 ($\pm$ 0.09) & 0.40 ($\pm$ 0.09) & 0.70 ($\pm$ 0.09) \\
Magnification  & 
$-4.04^{+0.12}_{-0.08}$ & $-3.18^{+ 0.14}_{-0.13}$ &  $-2.39^{+0.22}_{-0.23}$ & $-2.57^{+0.11}_{-0.07}$ \\
\hline
\hline
\end{tabular}}
\label{tab:params}
\end{table*}
\subsection{WFC3 photometry from subtracted images}
\label{ssec:phot_sub_joel}
On 2018 November 10, we obtained $HST$ WFC3 images of the lens and host galaxy system, long after the SN faded.
This allowed us to align the SN images and the template images and subtract the lens and host galaxy contribution (see Figure~\ref{fig:subtractions_rgb} for the combined RGB image, the template and the subtraction). This approach is freed from the assumptions of host and lens galaxy modeling. We note, however, that the challenge of lack of field stars and PSF sampling still remains.

On the subtracted images, we use four Moffat PSF models to simultaneously fit the SN images. The relative SN image positions are kept fixed. The errors are estimated by randomly putting small apertures on the residual image. For Images 1 and 2, the forward modelling (Sections~\ref{sec:wfc3_forwardmodel} and Appendix~\ref{sec:wfcpsf})  and template subtraction have very good agreement within the statistical errors. However, since Images 3 and 4 are significantly fainter, there are systematic differences in the template subtractions and the forward modelling. 
Examples of the SN data for 2016, October, 20, subtraction, model and residuals for all the WFC3 filters is shown in Figure~\ref{fig:subtractions}. Since this approach is significantly less model dependent, we use this version of the photometry in the analysis detailed below. The photometry is presented in section~\ref{app:phot_tables}.

\begin{figure*}
    \centering
    %\includegraphics[width=\textwidth, height=11cm, trim = 40 0 40 0 ]{hstdata_modelfit_indivimage_allfilt.pdf}
    %\vspace{.5mm}
    %\includegraphics[width=\textwidth, height=11cm, trim = 40 0 40 0]{grounddata_modelfit_summed_allfilt.pdf}
    \includegraphics[width=\textwidth]{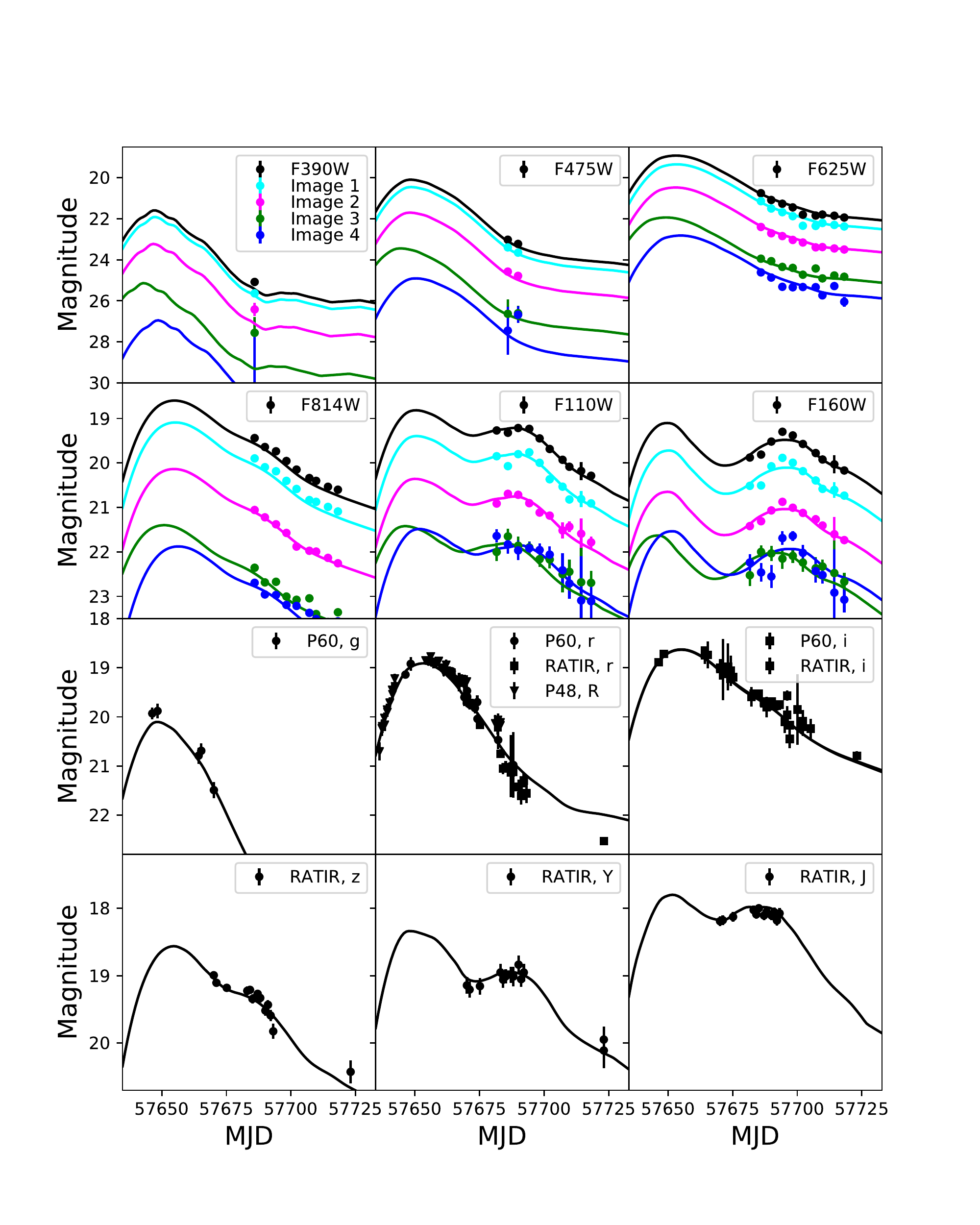}
    \caption{The multiple image model fit to the resolved photometry for the individual images from $HST$ (top six panels). The $HST$ data for Image 1 is in cyan, Image 2 in magenta, Image 3 in green and Image 4 in blue. The black lines show the combined data. The combined fit to the ground based data where the multiple images are not resolved (bottom six panels). 
      The filters are plotted in ascending order of effective wavelength. The RATIR, Y data are scaled down by 0.25 mag to account for unknown systematic effects.}
    \label{fig:16geu_lc}
\end{figure*}

\section{Lightcurve fitting model}
\label{sec:lcmodel}
In this section we present the model to fit the light curves of iPTF16geu.
We combine the ground based data with the resolved photometry from $HST$ to fit for the global lightcurve parameters: lightcurve shape (or stretch, $s$), color excess from either intrinsic color variations or dimming by dust in the host galaxy. In addition, we use the multi-band lightcurves of the resolved 
\sn images to fit for the lightcurve peaks and time offsets between the four \sn images, as well as extinction in the lensing galaxy for each individual line of sight. 
Here, we describe the SN model constructed to derive the time-delay, magnification and extinction parameters for iPTF16geu..
We construct a multiple-image model for an SN~Ia using \texttt{sncosmo} \citep{2016ascl.soft11017B} with a \texttt{Hsiao} model  for the SN~Ia spectral template, which is constructed from a large library of spectra for diverse SNe~Ia. The \texttt{Hsiao} model is appropriate since iPTF16geu shows light curve and spectroscopic properties consistent with normal SNe~Ia. 
SNe~Ia in the local universe show a characteristic near infrared (NIR; $iYJHK$) light curve morphology \citep{1996AJ....112.2438H,2010AJ....139..120F}.  Unlike in the optical, where the light curves decline after peak, in the NIR, SNe~Ia rebrighten 2-3 weeks after the $B$-band maximum. This feature has also been seen in well-studied intermediate-$z$ SNe \citep[e.g., see][]{2000ApJ...536...62R}. iPTF16geu shows a distinct second maxmium in the observer frame F110W and F160W filters. Accounting for time-dilation and K-correction, we find that the time of the second maximum ($t_2$)  29.3 $\pm$ 1.1 days after $B$-band maximum, consistent with the median value of $t_2$ for nearby, normal, SNe~Ia \citep{2012A&A...537A..57B, 2015MNRAS.448.1345D}. Since $t_2$ only depends on the redshift but not the distance, this further justifies the choice of using a normal SN~Ia model to fit the iPTF16geu light curves.

 We also correct the light curves for Milky Way (MW) dust. For all our dust corrections, we use the Milky Way reddening law \citep[][hereafter CCM89]{1989ApJ...345..245C}. In our fiducial analysis, we fix the $R_V$, for the CCM89 dust correction, to 2 in the host and lens galaxies and to the canonical value of 3.1 in the Milky Way. 
The value of host and lens galaxy $R_V$ is chosen from the slope of the luminosity-colour relation ($\beta \sim 3$, hence, $R_V \sim 2$) from the most updated cosmological samples of SNe~Ia \citep[see;][]{2018ApJ...859..101S,2018arXiv181102374D}. Since iPTF16geu shows light curve and spectroscopic features similar to core-normal SNe~Ia \citep[][Johansson et al. in prep]{2018MNRAS.473.4257C}, which are used for constraining cosmology, we can use the mean $R_V$ from the cosmological compilations when fitting for the light curve model parameters for iPTF16geu. Hence, in our model fit, we include the following parameters 
\begin{itemize}
    \item The stretch, $s$, for the SN
    \item The colour excess, $E(B-V)$ in the host galaxy
    \item The colour excess, $E(B-V)$ for the individual images in the lens galaxy 
    \item The time of maximum $t_{\rm max}$ for each of the four images, and hence, the time-delays between the images
    \item The magnification for each of the four images
\end{itemize}
\begin{equation}
\label{eq:chi}
\begin{aligned}
\chi ^2 &= \sum_{\lambda}^{\mathrm{Ground}}\frac{\left[\left(\sum_i^{N} {\rm F}_i^{\rm model}(t,\lambda)\right)-{\rm F}^{\rm data}(t,\lambda)\right]^2}{\sigma^2(t,\lambda)}+ \\
&\quad \sum_{\lambda}^{\mathrm{HST}}\sum_i^{N}\frac{\left[{\rm F}_i^{\rm model}(t,\lambda)-{\rm F}_i^{\rm data}(t,\lambda)\right]^2}{\sigma_i^2(t,\lambda)},
\end{aligned}
\end{equation}
We fit the model to the observations using a $\chi^2$ likelihood with two terms for the ground based and the $HST$ data. For the ground based data we compare the sum of the models to the observations, whereas for the $HST$ data we compare the individual images, where $F$ is the flux, $t$ is the epoch of the SN~Ia light curve, $N$ is the number of images and $\lambda$ is the effective wavelength of each filter. The model is fitted using a nested sampling software \texttt{nestle}\footnote{https://github.com/kbarbary/nestle} implemented in \texttt{sncosmo}.

\section{Results}
\label{sec:results}
In this section, we present the results of fitting the multiple-image SN model described in Section~\ref{sec:lcmodel} to the observations of iPTF16geu (see Figure~\ref{fig:16geu_lc}). In our analyses, we add an additional error term corresponding to 8$\%$ of the flux to the diagonal terms of the error covariance matrix for the $HST$ observations such that the reduced $\chi^2$ for the ground and space-based data individually (and hence, the  total) is  $\sim 1$. We note that the errors on the fitted parameters are severely underestimated without the additional error term to the likelihood, however, the best fit values are consistent with the values reported here. 
 We present the resulting values of the time-delays between the images (Section~\ref{ssec-time_delay}) and the properties of the extinction due to dust in the host and lens galaxies (Section~\ref{ssec-extinction}). For the fiducial case we fit the time of maximum, amplitudes and colour excesses for the four SN images as well as the total to selection absorption, and the colour excess in the host galaxy.

\subsection{Differential extinction and lensing magnification}
\label{ssec-extinction}
\begin{figure*}
    \centering
    \includegraphics[width=.48\textwidth]{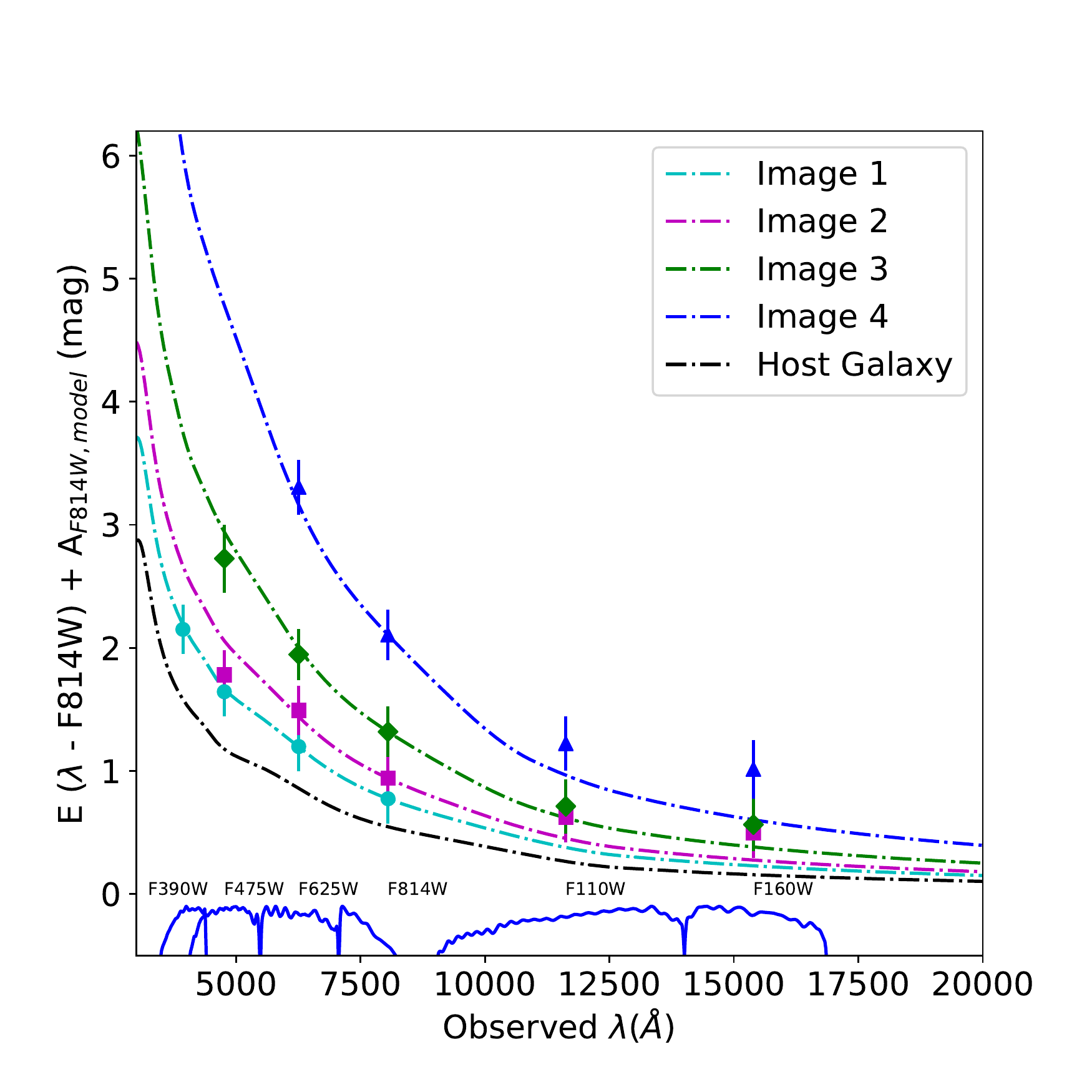}
    \includegraphics[width=.48\textwidth]{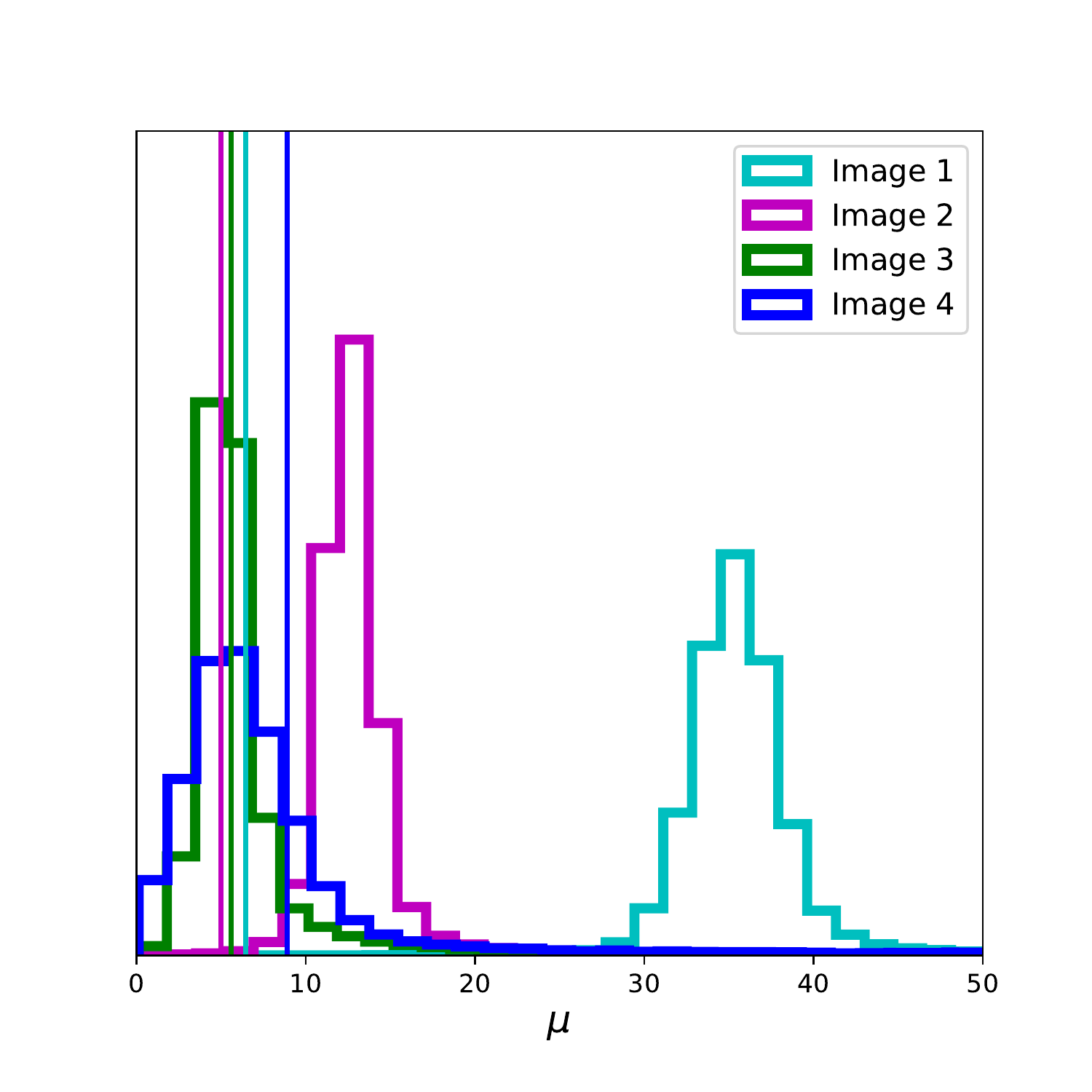}
    \caption{(Left) The observed colour excess for the resolved images in each filter relative to $F814W$ plus the model absorption in the $F814W$ filter compared to the best fit model absorption in each filter assuming the CCM89 dust law. The absorption from the host galaxy dust is plotted in black. For Image 1 we can see that the host galaxy is the dominant source of extinction, and for images 2,3,4 there is a progressively larger contribution from the dust in the lens galaxy.  (Right) magnification distribution for the individual images for the fiducial case of host and lens $R_V$ fixed to 2 compared to the predictions from the model assuming the lens to be a single isothermal ellipsoid (dashed-dotted lines; see M{\"o}rtsell et al in prep for details. The model prediction for $\mu$ of Image 2 has been shifted down by 0.5 so that it can be distinguished from the value for Image 3).}
    \label{fig:extinction_magnification}
\end{figure*}

Here, we present the properties of extinction of  iPTF16geu due to the dust in the host and the lens galaxies as well as the magnification of each image relative to a normal SNe~Ia at the redshift of the source (i.e. $z = 0.409$). As described in Section~\ref{sec:lcmodel}, for our fiducial case, we fix the $R_V$ in the host and lens galaxies to 2, the best fit value for SNe~Ia used in cosmology. 
The resulting parameters from the fit are summarised in Table~\ref{tab:params}. We assume that the SN host extinction is the same for each image since the difference in the light travel path is very small ($\sim$ 0.01 pc) and we do not expect the dust properties to vary significantly on those length scales. We find host $E(B-V)$ of 0.29 ($\pm 0.05$) mag. The lens $E(B-V)$ in the Images 1, 2, 3 and 4 is 0.06 ($\pm$ 0.08), 0.17 ($\pm$ 0.08), 0.42 ($\pm$ 0.09) and 0.94 ($\pm$ 0.07) mag respectively. The first two images have very little extinction in the lens galaxy, image 3 has moderate extinction, and image 4 is heavily extinguished. The combined host and lens galaxy absorption as a function of wavelength for the four images is shown in Figure~\ref{fig:extinction_magnification} (left panel).

\begin{figure}
    \centering
    \includegraphics[width=.5\textwidth]{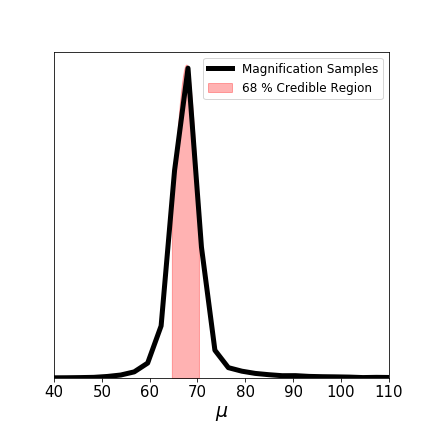}
    \caption{The posterior distribution for the total magnification fitting the multiple-image \texttt{sncosmo} model described in the text. The median value is 67.8 with a 68$\%$ credible region between (64.9, 70.3).}
    \label{fig:amp}
\end{figure}

We test the impact of altering the assumption on $R_V$. For a fixed $R_V = 3.1$, corresponding to the MW value, in both the host and lens galaxies, we find host $E(B-V)$ of 0.17 ($\pm$ 0.08) and lens $E(B-V)$ for Images 1, 2, 3 and 4 to be 0.13 ($\pm$ 0.08), 0.20 ($\pm$ 0.09), 0.40 ($\pm$ 0.09), 0.70 ($\pm$ 0.09). We also let the $R_V$ in the lens galaxy as a free parameter. The data indicate low lens $R_V < 1.8$ at 95 $\%$ C.L. Moreover, we fit for the $R_V$ for each individual line-of-sight to the four Images. We find that the limits on the $R_V$ at 95$\%$ C.L. are $<2.8$, $<3.2$, $<3.6$ and $<1.5$ for Images 1, 2, 3 and 4 respectively. We present details for the impact of dust properties on  the inferred magnification in section~\ref{app:indiv_rv}. 

We also compute the total magnification for iPTF16geu and the magnification of each image. For our fiducial case with both host and lens galaxy $R_V = 2$, we obtain a median magnification of 67.8 with a 68$\%$ credible region of (64.9, 70.3), plotted in Figure~\ref{fig:amp}. The distributions for the image magnifications are plotted in Figure~\ref{fig:extinction_magnification}. We test the impact of our assumption on the reddening law. We find that assuming a reddening law from \citet[hereafter F99]{Fitzpatrick:1999dx} doesn't change the total extinction for each image and hence, the inferred total magnification and magnification for each image is consistent with the value derived using the CCM89 dust law. We note, however, that the F99 dust law prefers a lower value of host $E(B-V)$ (and subsequently higher lens $E(B-V)$) than the fiducial case in Table~\ref{tab:params}.  
Since a higher host $E(B-V)$ is more consistent with spectroscopic observations (Johansson et al. in prep), we adopt the CCM89 dust law as our fiducial case. Furthermore, we find that the total magnification doesn't change when leaving both host and lens $R_V$ as free parameters and we get a similar constraint of host $R_V < 1.7$ at the 95$\%$ C.L.
 
Compared to the model predictions, which find Image 4 to be the brightest and the other three of similar brightness (M{\"o}rtsell et al. in prep), we find that the data suggest Image 1 is the brightest followed by Image 2 which is 0.44 times as bright and Images 3 and 4 which have similar brightness 0.2 and 0.26 times that of Image 1, after accounting for extinction corrections in each line of sight. The individual images are magnified by -3.88, -2.99, -2.19, -2.40 magnitudes respectively. We find that the magnification for iPTF16geu is robust to the assumption on the $R_V$ for the host and lens galaxies (see Table~\ref{tab:params} for individual magnifications for different $R_V$ assumptions). 
The individual image magnifications, using only the space based data are $-3.79^{+0.08}_{-0.07}$, $-2.90^{+0.10}_{-0.09}$, $-2.04^{+0.15}_{-0.15}$ and $-2.33^{+0.12}_{-0.16}$, consistent with the combined constraints.

 \subsection{Time-delays}
\label{ssec-time_delay}
We fit the time of maximum for the four images to the combination of the resolved $HST$ data and the unresolved ground based data (Equation~\ref{eq:chi}) and calculate the time-delay for images 2, 3 and 4 relative to Image 1, $t_{\rm max,2}$, $t_{\rm max,3}$, $t_{\rm max,4}$. The summary of the parameter values is presented in Table~\ref{tab:params}. The time-delays for the three images 2, 3 and 4 relative to Image 1 are -0.23, -1.43 and 1.35 days respectively. However, due to the large uncertainties in constraining the time-delays, they are consistent with a zero day time delay at the 95$\%$ C.L. The observed time-delays agree very well with the model predictions suggesting time-delays approximately one day \citep[][M{\"o}rtsell et al. in prep.]{2017ApJ...835L..25M}.
We also derive time-delays from the light curves without assuming an SN model. SNe~Ia have a distinct light curve morphology in the NIR, showing a second maximum $\sim$ few weeks after the first, we use the timing of this feature to derive constraints on $\Delta_t$. Since Images 1 and 2 are the brightest, the NIR light curves only have high enough signal to noise for these two images to derive a time-delay. We use two different methods to fit the data. Firstly, we derive $\Delta_{\rm t}$ using a Gaussian Process (GP) smoothing to the light curve. We use the Matern32 kernel implemented within the \texttt{GPy} package \citep{gpy2014}, and secondly a cubic spline interpolation. We obtain the timing of the second maximum for each image (and hence, the delay between the two) from the time at which the derivative is zero.
The model independent constraints are also consistent with the $\sim$ 1 day time delay derived from fitting the \texttt{sncosmo} model to the data. 
 Using only the $HST$ photometry, we get time-delays for images 2,3,4 of -0.57 ($\pm$ 1.13), -1.47 ($\pm$ 1.07), 0.79 ($\pm$ 1.23) days relative to image 1, consistent with the values from the combination of ground and space-based photometry.
\section{Anomalies between observed and predicted flux ratios}
\label{sec:substructure}
The fact that \geu exploded very close to the inner caustic of the lens makes the predicted radial position, magnification and arrival time very similar for all the images in a smooth lens model. 
In \citet{2017Sci...356..291G}, the magnification of the SN images could not be well constrained, but
the adopted lens halo predicted brightness differences between the SN images in disagreement with observations.
Taking advantage of the improved observational constraints since then, uncertainties in the mass, ellipticity and orientation of the lens galaxy are decreased by a factor $\sim 7$ (M{\"o}rtsell et al. in prep).

In table~\ref{tab:params}, we report the values of the magnifications for each individual image. We find that image 1 is the brightest and 4 is the faintest. Lens models of the system \citep[e.g.][M{\"o}rtsell et al. in prep.]{2017ApJ...835L..25M} suggest that the SN images would have very similar brightnesses since the SN images are symmetric around the lens. We test whether this could be a result of differential extinction, i.e. a difference in the $R_V$ for the dust in the region around Images 1 and 4. Assuming a different $R_V$ for images 1 and 4, in this case setting it to extreme values of $R_V = 1$ and $R_V = 5$ for images 1 and 4 respectively, we find that Image 4 is still fainter than image 1 by a factor $\sim$ 6 and the best fit value compensates for the high input $R_V$ with a lower inferred $E(B-V)$. Moreover, we also fitted for two different $R_V$'s in the lens, one for Images 1, 2 and 3 and a separate $R_V$ for Image 4. For this case, we get similar values for both $R_V$'s and hence, Image 4 is still $\sim$ 9 times fainter than image 1. We note also that deriving constraints assuming an F99 dust law also doesn't change the observed discrepancy between the brightness of images 1 and 4. Hence, differential extinction is an unlikely explanation for the discrepancy between the observed and modelled image brightness ratios. 

In an accompanying paper (M{\"o}rtsell et al. in prep), we present lens modelling for \geu and find that the preferred slope of the projected surface mass density is flatter than a single isothermal ellipsoid profile \citep{1994A&A...284..285K}, consistent with the observed time-delays. The observed fluxes cannot be explained by a smooth density profile, regardless of the slope and require a magnification of image 1 by  microlensing and a demagnification of images 2, 3 and 4. The differences between the observed fluxes and the smooth density profile are within the stellar microlensing predictions.  Discrepancies between the lens macromodel predictions and the observed flux ratios have previously been observed in multiply-imaged quasars, and attributed to microlensing \citep[e.g. in MG 0414+0534;][]{2018MNRAS.480.4675V}. However, out of the $\sim$ 200 known lensed quasar, only a few have a comparable angular separation to \geu \citep[see for e.g.][]{2018MNRAS.479.5060L,2019MNRAS.483.4242L}. Hence, since \geu probes high density regions near the core of the galaxy, it is not unusual that the macromodel predictions are discrepant with the observations.
While recent studies in the literature find that microlensing can add to the uncertainty in the measured time-delay \citep{2018ApJ...855...22G, 2019A&A...621A..55B}, these uncertainties are subdominant compared to the measurement error of $\sim$ 1 day error that we obtain from fitting the data. 
%\citet{2018ApJ...855...22G} find that the time-delay error due to microlensing is peaked at 1$\%$ for the colour curve 
%Recent studies in the literature study the impact of microlensing on the inferred time-delay, accounting for the 3D source geometry, and find a bias of a few tenths of a day \citep{2019A&A...621A..55B}. We note that the error on our inferred time-delays is $\sim 1$ day hence, the microlensing bias has a subdominant contribution to the error budget.

\section{Implications for observations of future strongly lensed \sn}
\label{sec:forecast}
For iPTF16geu, observations that resolved the multiple images only began $\sim$ 2 weeks after maximum light, hence, the light-curve peak in the $HST$ filters is not well determined. Here, we analyse what the constraints on the time-delay and extinction parameters would be if we were to obtain observations close to the peak. 
We use the best fit model for iPTF16geu to extrapolate the observations near maximum light. We generate ten observations uniformly in the phase region between 10 and 30 days from the first iPTF observations in the P48 R-band. We assume an error on each point to be drawn from a uniform distribution given by the errors on the observed post-maximum epochs. Since the SN is brighter at maximum light we would expect higher signal to noise for similar exposure times and a simpler PSF modeling leading to smaller errors on the fluxes. Under these assumptions we fit the above mentioned \texttt{sncosmo} model to the simulated data. 
We find that the maximum light data significantly improve the constraints on the time delays, with an error of 0.2 days ($\sim$ 5 hours) which is approximately four times lower than the error from the post-maximum observations. For the expected range of time-delays for ongoing and future surveys between 10 and 50 days \citep[see][for details]{2018arXiv180910147G},  this uncertainty corresponds to an error of 2$\%$ or lower in the time-delays, propagated into the $H_0$ uncertainty. Since typical model uncertainties in the lens modeling for quasars is $\sim 4\%$ \citep[e.g.][]{2019arXiv190702533C,2019arXiv190704869W}, and knowing the SNe~Ia magnification breaks the mass-sheet degeneracy, we can expect a typical $H_0$ uncertainty of 3-5$\%$ with the time-delay error having a subdominant contribution.
Moreover, studies in the literature \citep[e.g.][]{2018ApJ...855...22G} find that the impact of microlensing on time-delays is achromatic if the observations are obtained within approximately the first three weeks from explosion. The study finds that the time delay error from microlensing for glSNe in future surveys peaks at $1 \%$ for colour curve observations, hence, will be a subdominant contribution in the error budget. 
Therefore, near maximum observations are crucial for measuring $H_0$ precisely from time-delays, especially for highly symmetric systems which have short time-delays, like iPTF16geu.

\section{Conclusions}
\label{sec:conclusion}
In this paper we present ground-based and $HST$ follow-up of the first resolved, multiply-imaged gravitationally lensed SNe~Ia, iPTF16geu. Fitting a multiple image SN~Ia model to the data we were able to derive the total magnification, properties of extinction for each image and time-delays between the images. Accounting for the extinction in the individual images, we find that iPTF16geu is amplified by $67.8^{+2.6}_{-2.9}$ times relative to a normal SN~Ia at the redshift of the source. Since this value accounts for the extinction in each image separately, it is higher than the first estimate provided in G17. Assuming an $R_V =2$ extinction law in the host and lens galaxies, we find an $E(B-V) =0.06$ in Image 1 and 0.17 in Image 2, 0.41 mag in Image 3 and 0.94 mag in Image 4. 
From the multiple-image model fit, we find that time-delay of images 2, 3 and 4 relative to image 1 to be -0.23 ($\pm$ 0.99), -1.43 ($\pm$ 0.74) and 1.35 ($\pm$ 1.07) days, consistent with the limits presented in G17 and the model predictions in \citet{2017ApJ...835L..25M}. We use model independent smoothing techniques to derive the time-delay from the observations of the NIR second maximum and find consistent results with the multiple-image model fit. Furthermore, the total magnification and the time-delay estimates are robust to the assumptions on the host and lens $R_V$. 
We find that the observed difference in the brightness of Images 1 and 2 relative to images 3 and 4 is discrepant with model prediction which suggest a similar brightness for all images. Differential extinction is not sufficient to explain the observed discrepancy. This discrepancy can be possibly resolved with additional substructure lensing; further details are presented in an accompanying paper (M{\"o}rtsell et al. in prep). Finally, we presented forecasts for observations of lensed SNe discovered in the future and find that $\sim 10$ $HST$ observations of multiple-image glSNe~Ia around maximum light will improve the existing constraints on time-delays by a factor $\sim$ 4.

\section*{Acknowledgements}
We would like to thank Justin Pierel for interesting discussions on the time-delay computation. AG acknowledges support from the Swedish National Space Agency and the Swedish Research Council.

%%%%%%%%%%%%%%%%%%%%%%%%%%%%%%%%%%%%%%%%%%%%%%%%%%

%%%%%%%%%%%%%%%%%%%% REFERENCES %%%%%%%%%%%%%%%%%%

% The best way to enter references is to use BibTeX:

\bibliographystyle{mnras}
\bibliography{geu_followup} % if your bibtex file is called example.bib

%%%%%%%%%%%%%%%%%%%%%%%%%%%%%%%%%%%%%%%%%%%%%%%%%%

%%%%%%%%%%%%%%%%% APPENDICES %%%%%%%%%%%%%%%%%%%%%

\appendix
\section{Forward modelling of the \geu images}
\label{sec:bgmodel}
In this section we describe the forward model and fitting procedure for \geu images.
A model,  $F(r,\phi)$, of the observed 2D shape of the \geu system in a broadband image, can be expressed as a 
combination of parametric lens and host models, $L(r,\phi)$ and $H(r,\phi)$, and the point spread function (PSF) of the 
image as
\begin{eqnarray*}
	F(r,\phi) & = & A_n\cdot {\rm PSF}\otimes \left[L(r,\phi) + H(r,\phi)\right] +\\
	& + & \sum\limits_{i=1}^{4}\left( f_{i}^{(n)} {\rm PSF}(r_i,\phi_i) \right)+ B_n
\end{eqnarray*}
where the coordinates $(r,\phi)$ are defined with respect to the center (which are treated as nuisance parameters in the fit)
of the system in each image $n$, and the angle $0 \leq \phi < 2\pi$ runs from North towards East.  The index, $i=1,2,3,4$ runs 
over the four SN images, and $n$ runs over all observed epochs for a given band.  Furthermore, $f_{i}^{(n)}$ and $(r_i,\phi_i)$ are 
the fluxes and coordinates of the SN images.  The amplitude, $A_n$, can be used to account for varying photometric 
calibration, but must be kept fixed for at least one image in each band in order to break the degeneracy between the  
parameters for $L$ and $H$.  We also allow for the background, $B_n$ to vary between images. 

The lens, $L$, is modelled by a S\'ersic profile \citep{1963BAAA....6...41S}
\begin{equation}
	L(r,\phi) = S^{n_S}(r,\phi) = f_S\cdot \exp\left\{-b_n\left[\left(\frac{r_S(\theta,\epsilon;\phi)}{r_e}\right)^\frac{1}{2n_S} - 1\right]\right\}\,
	\label{eq:lens}
\end{equation}
where $r_S(\theta,\epsilon;\phi)$, is generalized to allow for an elliptical model with ellipticity, $\epsilon$, and rotation, $\theta$.  Here,
$b_n$ (solved for numerically) is defined such that $r_e$ contains half of the total luminosity , $f_S$ is the intensity, and $n_S$ is
the S\'ersic index.  
%In the cases where this model is not sufficient for describing the nucleus of the lensing galaxy we also add an 
%exponential, $S^1(r,\phi)$, with an effective radius, $r_\mathrm{exp}$, but with the same ellipticity and orientation 
%as for $S^{n_S}(r,\phi)$.

The host galaxy is expressed as a Gaussian profile according to
\begin{equation}
	H(x,y) = h\cdot f_H(\phi)\cdot\exp\left\{-\frac{(r - r_H(\phi))^2}{2\sigma_H^2}\right\}\, ,\label{eq:host}
\end{equation}
where  the amplitude, $f_H(\phi)$ and radius, $r_H(\phi)$ are defined as
\begin{eqnarray}
f_H(\phi) & = & \frac{a_0}{2} + \sum\limits_{j=1}^3 a_j\cdot\sin(j\phi) + b_j\cdot\cos(j\phi)\, ,\label{eq:hostflux} \\ 
r_H(\phi) & = & \frac{c_0}{2} +  c_1\cdot\sin(\phi) + d_1\cdot\cos(\phi)\, .\label{eq:hostradius}
% \sigma_H(\phi) & = & \frac{g_0}{2} + \sum\limits_{j=1}^3 g_j\cdot\sin(j\phi) + h_j\cdot\cos(j\phi)\, .
\end{eqnarray}

\begin{table}
	\centering
	\caption{%
		The fitted SN positions to the NIRC2 \jband-band images.  The origin is defined as the origin of the \geu system 
		as it is obtained from the model fit and the angle, $0\leq\varphi<2\pi$, is defined from North towards East. All 
		quoted uncertainties are statistical errors obtained from the simultaneous fit described in the text.}
		\label{tb:snpos}
	\begin{tabular}{c r@{\hspace{0.5em}}l r@{\hspace{0.5em}}l}
\hline\hline
Image & \multicolumn{2}{c}{$r$} & \multicolumn{2}{c}{$\varphi$}\\
      & \multicolumn{2}{c}{$('')$} & \multicolumn{2}{c}{(rad)}\\
\hline
1 & 0.251 & (0.001)& 4.468 & (0.002)\\
2 & 0.324 & (0.001)& 2.679 & (0.003)\\
3 & 0.297 & (0.002)& 1.013 & (0.006)\\
4 & 0.276 & (0.001)& 5.860 & (0.005)\\
\hline\hline
\end{tabular}

\end{table}

%% TABLE: Lens model parameters
\begin{table*}
	\centering
	\caption{%
		The fitted lens model parameters as described in eq.~\eqref{eq:lens}.  The origin $(x_0^{(n)},y_0^{(n)})$ of the coordinate
		system are fitted as free parameters for each image $n$.  Angles are defined as $0\leq\theta<2\pi$ from North towards East.
		All quoted uncertainties are statistical fitting errors from each simultaneous fit.
		{\em Note:} Parameters marked with an asterisk ($^{*}$) were fixed to the given value in the fit.}
		\label{tb:lensmodel}
	\begin{tabular}{l | r @{\hspace{0.5em}} l r @{\hspace{0.5em}} l r @{\hspace{0.5em}} l r @{\hspace{0.5em}} l r @{\hspace{0.5em}} l}
\hline\hline
Filter & \multicolumn{2}{c}{$f_S$} & \multicolumn{2}{c}{$r_e$} & \multicolumn{2}{c}{$n_S$} & \multicolumn{2}{c}{$\varepsilon$} & \multicolumn{2}{c}{$\theta$}\\
  & \multicolumn{2}{c}{} & \multicolumn{2}{c}{$('')$} & \multicolumn{2}{c}{ } & \multicolumn{2}{c}{ } & \multicolumn{2}{c}{(rad)}\\
\hline
$K_s$ & 2.43E$-$01 & (1E$-$03) & 0.544 & (0.001) & 0.79 & (0.00) & 0.165 & (0.002) & 0.35 & (0.01)\\
$H$ & 2.08E$-$01 & (2E$-$03) & 0.585 & (0.002) & 0.84 & (0.01) & 0.160 & (0.002) & 0.30 & (0.01)\\
$F160W$ & 1.01E$+$01 & (4E$-$01) & 0.532 & (0.010) & 1.57 & (0.04) & $^*$0.160 &  & $^*$0.30 & \\
$J$ & 1.44E$-$01 & (1E$-$03) & 0.528 & (0.002) & 0.79 & (0.01) & 0.127 & (0.002) & 0.19 & (0.01)\\
$F110W$ & 1.52E$+$01 & (7E$-$01) & 0.529 & (0.013) & 1.66 & (0.05) & $^*$0.127 &  & $^*$0.19 & \\
$F814W$ & 1.45E$-$01 & (9E$-$03) & 0.619 & (0.026) & 1.56 & (0.06) & 0.252 & (0.005) & 0.22 & (0.01)\\
$F625W$ & 1.16E$-$01 & (1E$-$03) & 0.518 & (0.006) & $^*$1.56 &  & $^*$0.252 &  & $^*$0.22 & \\
$F475W$ & 3.69E$-$02 & (2E$-$03) & 0.474 & (0.031) & $^*$1.56 &  & $^*$0.252 &  & $^*$0.22 & \\
$F390W$ & 6.13E$-$03 & (1E$-$03) & 0.588 & (0.147) & $^*$1.56 &  & $^*$0.252 &  & $^*$0.22 & \\
\hline\hline
\end{tabular}

\end{table*}

%% TABLE: Host amplitude and radius parameters
\begin{table*}
	\centering
	\caption{%
		The fitted parameters of the host model, defined in~\eqref{eq:host}.  All quoted uncertainties are statistical 
		fitting errors from each simultaneous fit.
		{\em Note:} Parameters marked with an asterisk ($^{*}$) were fixed to the given value in the fit.}
		\label{tb:hostmodelflux}
	\begin{tabular}{l | r @{\hspace{0.5em}} l r @{\hspace{0.5em}} l r @{\hspace{0.5em}} l r @{\hspace{0.5em}} l r @{\hspace{0.5em}} l}
\hline\hline
Filter & \multicolumn{2}{c}{$a_0$} & \multicolumn{2}{c}{$a_1$} & \multicolumn{2}{c}{$b_1$} & \multicolumn{2}{c}{$a_2$} & \multicolumn{2}{c}{$b_2$}\\
  & \multicolumn{2}{c}{} & \multicolumn{2}{c}{} & \multicolumn{2}{c}{} & \multicolumn{2}{c}{} & \multicolumn{2}{c}{}\\
\hline
$K_s$ & 1.47E$+$00 & (1E$-$02) & $-$2.97E$-$01 & (3E$-$03) & 2.56E$-$01 & (3E$-$03) & $-$1.41E$-$01 & (2E$-$03) & 3.53E$-$02 & (2E$-$03)\\
$H$ & 5.16E$-$01 & (3E$-$03) & $-$1.07E$-$01 & (2E$-$03) & 8.64E$-$02 & (2E$-$03) & $-$4.76E$-$02 & (1E$-$03) & 1.75E$-$03 & (2E$-$03)\\
$F160W$ & 1.01E$+$02 & (1E$+$00) & $-$1.48E$+$01 & (7E$-$01) & 2.33E$+$01 & (7E$-$01) & $-$2.39E$+$01 & (6E$-$01) & 4.88E$+$00 & (6E$-$01)\\
$J$ & 2.78E$-$01 & (5E$-$03) & $-$6.93E$-$02 & (2E$-$03) & 4.75E$-$02 & (2E$-$03) & $-$3.43E$-$02 & (1E$-$03) & $-$1.52E$-$02 & (1E$-$03)\\
$F110W$ & 1.40E$+$02 & (2E$+$00) & $-$2.28E$+$01 & (1E$+$00) & 2.62E$+$01 & (1E$+$00) & $-$3.85E$+$01 & (1E$+$00) & $-$2.03E$+$00 & (1E$+$00)\\
$F814W$ & 1.41E$+$00 & (3E$-$02) & $-$1.11E$-$01 & (8E$-$03) & 1.15E$-$01 & (8E$-$03) & $-$1.51E$-$01 & (9E$-$03) & $-$8.07E$-$02 & (9E$-$03)\\
$F625W$ & 8.06E$-$01 & (4E$-$02) & 4.52E$-$02 & (7E$-$03) & 2.73E$-$02 & (7E$-$03) & $-$9.13E$-$02 & (9E$-$03) & $-$4.77E$-$02 & (8E$-$03)\\
$F475W$ & 1.07E$-$01 & (2E$-$02) & $^*$0.00E$+$00 &  & $^*$0.00E$+$00 &  & $^*$0.00E$+$00 &  & $^*$0.00E$+$00 & \\
\hline
Filter & \multicolumn{2}{c}{$a_3$} & \multicolumn{2}{c}{$b_3$} & \multicolumn{2}{c}{$c_0$} & \multicolumn{2}{c}{$c_1$} & \multicolumn{2}{c}{$d_1$}\\
  & \multicolumn{2}{c}{} & \multicolumn{2}{c}{} & \multicolumn{2}{c}{$('')$} & \multicolumn{2}{c}{$('')$} & \multicolumn{2}{c}{$('')$}\\
\hline
$K_s$ & 1.76E$-$01 & (2E$-$03) & $-$1.63E$-$01 & (3E$-$03) & 5.831E$-$01 & (3E$-$04) & 3.08E$-$02 & (3E$-$04) & $-$3.54E$-$02 & (3E$-$04)\\
$H$ & 5.72E$-$02 & (2E$-$03) & $-$4.73E$-$02 & (2E$-$03) & 5.787E$-$01 & (8E$-$04) & 3.38E$-$02 & (8E$-$04) & $-$3.11E$-$02 & (8E$-$04)\\
$F160W$ & $-$1.15E$+$00 & (8E$-$01) & $-$1.26E$+$01 & (6E$-$01) & $^*$5.787E$-$01 &  & $^*$3.38E$-$02 &  & $^*$$-$3.11E$-$02 & \\
$J$ & 3.46E$-$02 & (1E$-$03) & $-$2.00E$-$02 & (1E$-$03) & 5.911E$-$01 & (1E$-$03) & 4.49E$-$02 & (1E$-$03) & $-$1.69E$-$02 & (9E$-$04)\\
$F110W$ & $-$4.29E$-$01 & (2E$+$00) & $-$1.54E$+$01 & (1E$+$00) & $^*$5.911E$-$01 &  & $^*$4.49E$-$02 &  & $^*$$-$1.69E$-$02 & \\
$F814W$ & 9.74E$-$02 & (9E$-$03) & $-$1.71E$-$01 & (9E$-$03) & 5.784E$-$01 & (1E$-$03) & 3.51E$-$02 & (8E$-$04) & $-$6.58E$-$03 & (8E$-$04)\\
$F625W$ & 1.74E$-$02 & (9E$-$03) & $-$1.13E$-$01 & (9E$-$03) & 5.788E$-$01 & (2E$-$03) & 3.28E$-$02 & (1E$-$03) & 5.30E$-$03 & (1E$-$03)\\
$F475W$ & $^*$0.00E$+$00 &  & $^*$0.00E$+$00 &  & $^*$5.788E$-$01 &  & $^*$3.28E$-$02 &  & $^*$5.30E$-$03 & \\
\hline\hline
\end{tabular}

\end{table*}
The parameter $h$, and the parameters $a_0, a_j, b_j$ cannot be fitted simultaneously.  The former is only allowed as a free parameter
for \wfcir when $a_0, a_j, b_j$ is fixed to the solution for the LGS-AO data.
We use the parametrisation by \citet{1969A&A.....3..455M} for the PSF.  While it is customary to use isolated bright stars in the field to fit the PSF shape,
and then fix the model, we are lacking 
isolated stars (or any objects but the \geu system) in the narrow field of view NIRC2 images. Hence, the PSF parameters is fitted 
together with the rest of the parameters of the \geu model, including the \sn fluxes.  In other words, the \sn images is effectively 
used to determine the PSF shape.  We do not allow the PSF shape parameters to vary between the four \sn images.
Fitting the PSF shape together with the model will complicate the fitting procedure.  While the \geu model is not expected to vary with
time, the observing conditions will, and these are characterised by the differences in the PSFs we are attempting to fit together 
with the model.  We address this by iteratively fitting first the model, and then the PSF shape, to one of our epochs.  Once this fit has
converged, we use the resulting parameters as initial conditions for the simultaneous fit.  For each iteration in the final 
fitting procedure, the combined model then first convolved with the PSF before it is compared to the data.   For the reference 
images, that are lacking point-sources, we adopt a Gaussian PSF.  The width of the Gaussian is fitted together with the 
other parameters, and the degeneracy between the Gaussian width and and the background model is broken since the same 
background model is fitted to all epochs simultaneously. 

We also allow for a rotation, $\delta_n$ of the full system, i.e. $\phi \rightarrow \phi + \delta_n$ between images.  These parameters 
must be fixed for at least one image to break the degeneracy with the angle dependent model parameters.

In its most general implementation the number of free parameters of the model can be:
\begin{itemize}
\item $2\times n + (n-1)$ for the position and rotation in each image
\item $4\times 3 \times n$ for the SN fluxes and positions
\item $n-1$ for the normalization of the model in each image
\item $5$ for the lens model
\item $11$ for the host model
\end{itemize}
which results in a total of $14 + 16n$ parameters.  However, when fitting this model to the data we will generally require 
that the SN positions are the same between different filters.  
%In practice, we also applied an iterative approach where the complexity of the model was increased in each step and the Bayesian 
%Information Criterion \citep[BIC,][]{schwarz1978} was calculated.  The more complex model was only kept if  $\Delta\mathrm{BIC}>10$.
%% TABLE: SN image positions

%% WFC3 PSF model
%%
\section{Empirical point-spread function model for \wfc}
\label{sec:wfcpsf}
Here we describe the point-spread function (PSF) model for the \wfc data for iPTF16geu.
Although the shape and variability of the PSF of \wfc has been studied in great detail 
\citep{2016wfc..rept...12A,2017arXiv170600386A}, a simple time-independent PSF model will be used here.  
We fit the profile,
\[
	PSF(x,y;A,\alpha,\gamma) = A\cdot \frac{\alpha - 1}{\pi\gamma^2}\left[1 + \left(\frac{x^2 + y^2}{\gamma^2}\right)\right]^{-\alpha}\,,
\]
as described by \citet{1969A&A.....3..455M},  where, $x$ and $y$ are pixel coordinates and $A$, $\alpha$ and $\gamma$ are 
free parameters.  The profile is fitted to bright isolated stars observed between 2010 and 2017 and with the same subarray, 
\uvisaperture, used for the \geu.  The data are also drizzled to the same resolution, and with the same kernel as for our science 
observations.

For the \wfcir data it is more challenging to fit the full model due to the lower resolution, the broader PSF, and the lower flux ratio 
between the \sn images and the background.   The effective radius and the S\'ersic index for the lens can be constrained by the 
light beyond the Einstein ring, but there is a degeneracy between the different components inside the ring.  Here, we fixed the 
$\varepsilon$ and $\theta$ parameters of the lens model for \hstj and \hsth to the corresponding NIRC2 results for the 
\jband and \hband, respectively, since the effective wavelengths of these filters are similar.  Similarly, the radial dependence and the width
of the host model were also fixed to the corresponding NIRC2 results.

\begin{figure}
\centering
\includegraphics[width=\columnwidth]{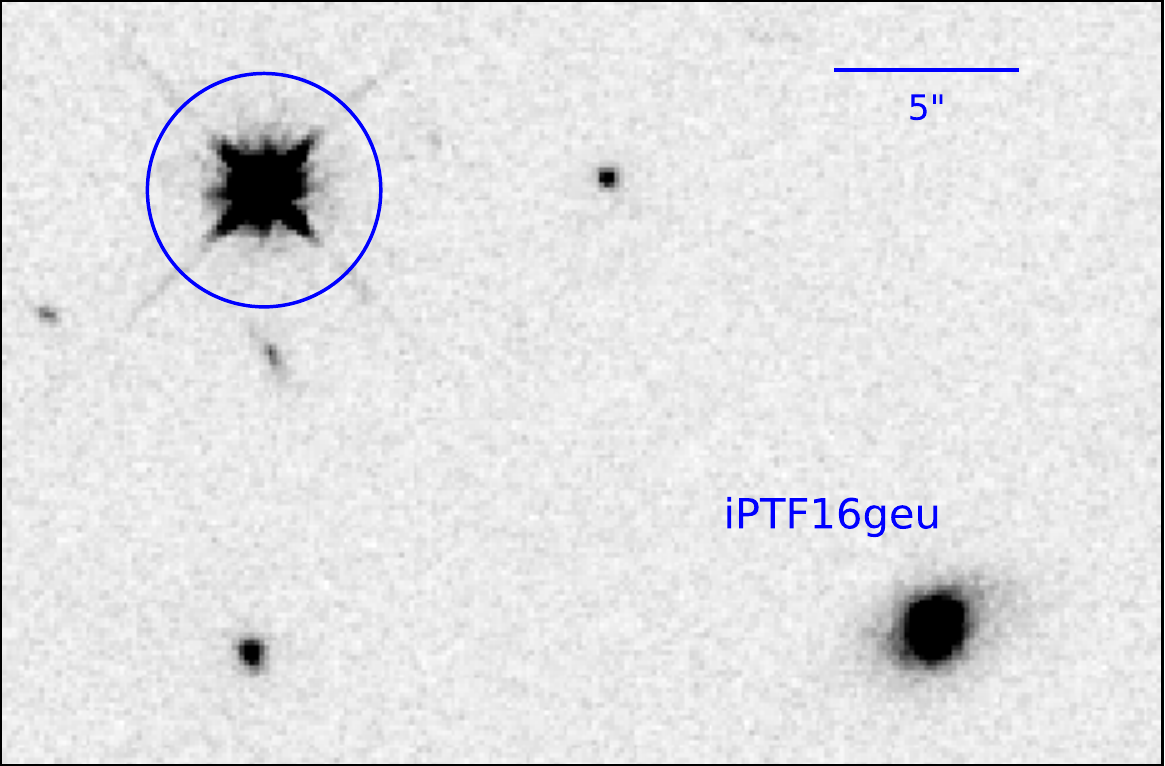}
\caption{%
	Observation of \geu in the F110W band obtained on 2016, November 17.  The blue circle marks the bright star visible in
	all IR observations, that was used for determining the aperture correction.
	\label{fig:psfnir}
}
\end{figure}

The \wfcir observations of \geu was obtained with a larger field-of-view than \wfcuvis as shown in Fig.~\ref{fig:psfnir}, and 
included a bright, but unsaturated, star $\sim20\arcsec$ North of the object.  The star was visible in all \wfcir and bands and
could be used to determine the PSF shape for these filters using the same approach as for \wfcuvis. 
%% For the WFC3/IR observations of iPTF16geu we offset the pointing with about 20" (160 pixels) from the center to avoid having
%% the nearby bright star to fall on the detector.  On the other hand the, for the observations of standard stars used for 
%% deriving the PSF the pointing was never chosen to that the star fell in the center. All standard star observations were obtained using
%% smaller 256x256 subarray.

The PSF fit is carried out to all data in a given filter.  While the amplitude, $A$, is fitted to each exposure, the parameters, 
$\alpha$ and $\gamma$, are only allowed to vary between different filters.   The fitted amplitudes of each star are then compared to the 
corresponding value obtained from aperture photometry.  For the latter we follow the guidelines in the \wfc data handbook
\citep{wfc3handbook} and always use a fixed aperture radius of $0.4\arcsec$, for which zero points have been derived.  
The two sets of values are used both to calculate the aperture correction for the PSF photometry and to assess the quality of this simple 
time-independent PSF model.  The latter is quantified by the estimated standard deviation between the PSF and and aperture
photometry for all observations in a given filter.  These values are then added in quadrature to the photometry error budget.  

Since we assume that the PSF is the same for all epochs, we do not convolve the model before comparing with the data in the fitting routine, although doing this will result in similar results.  For $F814W$, we fit the same parameters as described in \S\ref{sec:nirc2}.
For the remaining \wfcuvis filters this was not possible, and the lens parameters $n_S$, $\varepsilon$ and $\theta$ were then fixed to the
results obtained for $F814W$.  For \hstr, we fit all host parameters, while a simplified model with only the first order of angular dependence 
was used for \hstb.  Furthermore, the radial dependence and the width of the host model was fixed to the results obtained for $F814W$.   For \hstu, the host component was omitted completed.  The model complexity for each filter was determined by calculating the  impact on the Bayesian Information Criteria (BIC) as more parameters were added.  When additional parameters did not impact the BIC, the previous model was selected.

The fitted PSF parameters and the $\sigma$ values are given in Table~\ref{tb:psf} for the different filters.  For the \wfcir data the last 
epoch from 2016, Nov 22 was excluded from the fits since this was found to deviate significantly from the others.
% and a  couple of examples of profile fits are shown in Fig.~\ref{fig:psf}.

\begin{table}
\centering
\caption{%
   Table of the derived PSF parameters and the estimated standard deviations PSF and aperture photometry 
   for all stars in each given.  Here, the full width at half maximum was calculated as 
   $\mathrm{FWHM} = 2\gamma\left(2^{1/\alpha}-1\right)^{1/2}$. See the text for further details.
  \label{tb:psf}
}
\begin{tabular}{ccccc}
\hline\hline
Band & $\alpha$ & $\gamma$ & FWHM & $\sigma$ \\
  & & (pix) & (pix) & (mag)\\
\hline
F390W & 3.245 & 2.398 & 2.340 & 0.072 \\
F475W & 3.114 & 2.531 & 2.528 & 0.096 \\
F625W & 1.844 & 1.592 & 2.151 & 0.083 \\
F814W & 1.798 & 1.414 & 1.939 & 0.024 \\
F110W & 2.929 & 1.343 & 1.388 & 0.007 \\
F160W & 2.646 & 1.327 & 1.452 & 0.006 \\
\hline\hline
\end{tabular}

\end{table}

\section{Photometry Tables for \geu}
\label{app:phot_tables}
In this section we present the photometry for each of the four resolved images of \geu as well as the unresolved data from ground-based facilities used for the analyses in this paper.

\begin{landscape}
\begin{table}
	%\centering
	\caption{%
		Photometry for the individual resolved images derived from image subtractions using the $HST$ references  (see Section~\ref{ssec:phot_sub_joel} for details)
	\label{tb:phot_sub_joel}}
	\begin{tabular}{cccccccccc}
\hline\hline
MJD & Filter & Flux 1 & $\sigma$ & Flux 2 & $\sigma$ & Flux 3 & $\sigma$ & Flux 4 & $\sigma$\\
\multicolumn{1}{c}{(days)} & & \multicolumn{2}{c}{(counts/s)} & \multicolumn{2}{c}{(counts/s)} & \multicolumn{2}{c}{(counts/s)} & \multicolumn{2}{c}{(counts/s)} \\
\hline
57685.952 & F390W & 0.724 & 0.126 & 0.352 & 0.126 & 0.123 & 0.126 & 0.01 & 0.126 \\
57685.949 & F475W & 7.639 & 0.357 & 2.578 & 0.317 & 0.387 & 0.355 & 0.182 & 0.353 \\
57689.921 & F475W & 6.034 & 0.183 & 2.128 & 0.165 & 0.397 & 0.16 & 0.371 & 0.167 \\
57685.93 & F625W & 51.713 & 0.36 & 16.248 & 0.319 & 3.908 & 0.311 & 2.112 & 0.314 \\
57689.923 & F625W & 37.214 & 0.258 & 12.243 & 0.231 & 3.499 & 0.231 & 1.674 & 0.229 \\
57694.217 & F625W & 31.531 & 0.141 & 10.781 & 0.127 & 2.702 & 0.125 & 1.105 & 0.124 \\
57698.255 & F625W & 26.375 & 0.16 & 9.073 & 0.142 & 2.586 & 0.14 & 1.082 & 0.139 \\
57702.166 & F625W & 17.133 & 0.181 & 8.084 & 0.168 & 1.904 & 0.164 & 1.094 & 0.166 \\
57707.131 & F625W & 17.013 & 0.153 & 6.499 & 0.136 & 2.518 & 0.135 & 1.089 & 0.135 \\
57709.731 & F625W & 19.432 & 0.133 & 6.586 & 0.121 & 1.613 & 0.119 & 0.753 & 0.12 \\
57714.343 & F625W & 17.807 & 0.153 & 6.188 & 0.14 & 1.838 & 0.138 & 1.142 & 0.14 \\
57718.186 & F625W & 16.624 & 0.151 & 5.907 & 0.141 & 1.737 & 0.139 & 0.564 & 0.141 \\
57685.947 & F814W & 109.684 & 0.621 & 37.694 & 0.541 & 11.404 & 0.532 & 8.352 & 0.543 \\
57689.925 & F814W & 91.459 & 0.363 & 32.381 & 0.306 & 8.419 & 0.301 & 6.558 & 0.305 \\
57694.219 & F814W & 84.043 & 0.239 & 28.041 & 0.2 & 8.525 & 0.197 & 6.485 & 0.199 \\
57698.257 & F814W & 68.913 & 0.294 & 23.438 & 0.259 & 6.282 & 0.256 & 5.282 & 0.257 \\
57702.168 & F814W & 58.245 & 0.239 & 17.679 & 0.223 & 5.878 & 0.221 & 5.173 & 0.223 \\
57707.133 & F814W & 46.287 & 0.299 & 16.195 & 0.284 & 6.057 & 0.283 & 4.481 & 0.284 \\
57709.763 & F814W & 44.643 & 0.276 & 15.938 & 0.263 & 4.374 & 0.261 & 3.945 & 0.262 \\
57714.345 & F814W & 40.195 & 0.26 & 13.987 & 0.235 & 3.712 & 0.232 & 3.043 & 0.233 \\
57718.189 & F814W & 36.566 & 0.253 & 12.51 & 0.234 & 4.537 & 0.233 & 3.763 & 0.235 \\
57681.641 & F110W & 614.912 & 17.284 & 230.826 & 17.368 & 84.698 & 17.163 & 118.027 & 17.329 \\
57685.99 & F110W & 500.802 & 20.789 & 282.373 & 21.0 & 116.957 & 20.809 & 99.238 & 21.027 \\
57689.965 & F110W & 641.963 & 19.965 & 276.658 & 19.975 & 96.479 & 19.72 & 87.506 & 19.761 \\
57694.263 & F110W & 667.005 & 12.518 & 232.676 & 12.572 & 93.027 & 12.386 & 91.516 & 12.498 \\
57698.266 & F110W & 536.644 & 13.021 & 191.472 & 13.091 & 73.037 & 12.926 & 88.226 & 13.063 \\
57702.175 & F110W & 381.608 & 15.014 & 179.868 & 15.073 & 72.463 & 14.965 & 80.242 & 15.051 \\
57707.142 & F110W & 327.289 & 24.493 & 132.207 & 24.504 & 53.472 & 24.458 & 58.055 & 24.485 \\
57709.785 & F110W & 251.143 & 16.229 & 142.796 & 16.249 & 56.224 & 16.174 & 44.146 & 16.208 \\
57714.354 & F110W & 253.175 & 46.9 & 123.602 & 46.909 & 45.288 & 46.89 & 31.003 & 46.901 \\
57718.196 & F110W & 231.738 & 12.747 & 103.494 & 12.763 & 44.747 & 12.702 & 30.562 & 12.731 \\
57681.644 & F160W & 148.62 & 5.715 & 64.607 & 5.728 & 23.347 & 5.713 & 30.422 & 5.734 \\
57685.992 & F160W & 149.738 & 5.307 & 71.464 & 5.327 & 37.903 & 5.27 & 24.748 & 5.318 \\
57689.968 & F160W & 222.888 & 6.139 & 89.367 & 6.158 & 36.708 & 6.108 & 22.744 & 6.145 \\
57694.265 & F160W & 265.299 & 7.666 & 106.757 & 7.685 & 32.91 & 7.625 & 50.389 & 7.676 \\
57698.269 & F160W & 239.893 & 5.68 & 94.505 & 5.689 & 34.862 & 5.588 & 52.747 & 5.654 \\
57702.177 & F160W & 201.365 & 6.465 & 84.814 & 6.48 & 30.358 & 6.462 & 37.136 & 6.485 \\
57707.145 & F160W & 166.162 & 6.944 & 74.411 & 6.952 & 27.0 & 6.917 & 25.435 & 6.938 \\
57709.787 & F160W & 139.532 & 4.371 & 65.376 & 4.395 & 27.973 & 4.363 & 23.446 & 4.402 \\
57714.357 & F160W & 136.585 & 23.648 & 54.621 & 23.651 & 24.437 & 23.651 & 16.295 & 23.655 \\
57718.198 & F160W & 121.397 & 4.181 & 48.36 & 4.193 & 20.446 & 4.166 & 14.082 & 4.189 \\
\hline
\end{tabular}

\end{table}
\end{landscape}

\begin{landscape}
\begin{table}
	%\centering
	\caption{%
		Ground based photometry from Palomar P48 and P60 and RATIR used in the analyses.
	\label{tb:phot_ground}}
	\resizebox{.3\textwidth}{!}{\begin{tabular}{ccccHHH}
\hline\hline
MJD & Filter & Flux & $\sigma$ & ZP & zpsys & imageid \\
\multicolumn{1}{c}{(days)} & & \multicolumn{2}{c}{(counts/s)} &&  \\
\hline
57636.335 & P48R & 52.481 & 9.6673 & 25.0 & ab & 0\\
57637.181 & P48R & 80.91 & 12.668 & 25.0 & ab & 0\\
57637.329 & P48R & 83.946 & 11.598 & 25.0 & ab & 0\\
57638.177 & P48R & 98.175 & 10.851 & 25.0 & ab & 0 \\
57638.328 & P48R & 86.298 & 11.128 & 25.0 & ab & 0 \\
57639.227 & P48R & 114.82 & 13.747 & 25.0 & ab & 0 \\
57639.337 & P48R & 109.65 & 10.099 & 25.0 & ab & 0 \\
57640.215 & P48R & 130.62 & 13.233 & 25.0 & ab & 0 \\
57640.324 & P48R & 130.62 & 13.233 & 25.0 & ab & 0 \\
57641.201 & P48R & 165.96 & 18.342 & 25.0 & ab & 0 \\
57641.318 & P48R & 152.76 & 12.662 & 25.0 & ab & 0 \\
57642.196 & P48R & 205.12 & 18.892 & 25.0 & ab & 0 \\
57642.31 & P48R & 177.01 & 17.934 & 25.0 & ab & 0 \\
57646.329 & P60r & 220.8 & 16.269 & 25.0 & ab & 0 \\
57646.331 & P60i & 277.97 & 10.241 & 25.0 & ab & 0 \\
57646.333 & P60g & 106.66 & 12.771 & 25.0 & ab & 0 \\
57648.359 & P60r & 270.4 & 37.357 & 25.0 & ab & 0 \\
57648.361 & P60i & 325.09 & 17.965 & 25.0 & ab & 0 \\
57648.364 & P60g & 111.69 & 16.459 & 25.0 & ab & 0 \\
57654.14 & P48R & 288.4 & 18.594 & 25.0 & ab & 0 \\
57655.115 & P48R & 283.14 & 23.47 & 25.0 & ab & 0 \\
57655.24 & P48R & 283.14 & 18.255 & 25.0 & ab & 0 \\
57656.124 & P48R & 310.46 & 22.875 & 25.0 & ab & 0 \\
57656.25 & P48R & 277.97 & 20.482 & 25.0 & ab & 0 \\
57657.148 & P48R & 272.9 & 17.594 & 25.0 & ab & 0 \\
57657.246 & P48R & 270.4 & 24.904 & 25.0 & ab & 0 \\
57658.132 & P48R & 280.54 & 20.671 & 25.0 & ab & 0 \\
57658.175 & P48R & 272.9 & 17.594 & 25.0 & ab & 0 \\
57659.245 & P48R & 285.76 & 23.687 & 25.0 & ab & 0 \\
57660.149 & P48R & 251.19 & 20.822 & 25.0 & ab & 0 \\
57661.109 & P48R & 246.6 & 20.442 & 25.0 & ab & 0 \\
57661.154 & P48R & 231.21 & 19.165 & 25.0 & ab & 0 \\
57662.104 & P48R & 265.46 & 26.895 & 25.0 & ab & 0 \\
57662.228 & P48R & 237.68 & 17.513 & 25.0 & ab & 0 \\
57663.119 & P48R & 226.99 & 16.725 & 25.0 & ab & 0 \\
57663.222 & P48R & 224.91 & 18.643 & 25.0 & ab & 0 \\
57664.17 & P48R & 229.09 & 18.99 & 25.0 & ab & 0 \\
57664.249 & P60r & 222.84 & 18.472 & 25.0 & ab & 0 \\
57664.251 & P60i & 346.74 & 28.742 & 25.0 & ab & 0 \\
57664.261 & P60r & 235.5 & 13.014 & 25.0 & ab & 0 \\
57664.263 & P60i & 316.23 & 55.339 & 25.0 & ab & 0 \\
57664.264 & P60g & 48.306 & 8.0084 & 25.0 & ab & 0 \\
57665.294 & P60r & 205.12 & 18.892 & 25.0 & ab & 0 \\
57665.296 & P60i & 319.15 & 91.125 & 25.0 & ab & 0 \\
57665.298 & P60g & 52.966 & 7.8054 & 25.0 & ab & 0 \\
57667.106 & P60r & 192.31 & 37.196 & 25.0 & ab & 0 \\
57668.099 & P48R & 203.24 & 18.719 & 25.0 & ab & 0 \\
57668.216 & P48R & 190.55 & 17.55 & 25.0 & ab & 0 \\
57669.113 & P48R & 180.3 & 23.249 & 25.0 & ab & 0 \\
\hline
\end{tabular}}
\resizebox{.3\textwidth}{!}{\begin{tabular}{ccccHHH}
\hline\hline
MJD & Filter & Flux & $\sigma$ & ZP & zpsys & imageid \\
\multicolumn{1}{c}{(days)} & & \multicolumn{2}{c}{(counts/s)} &&  \\
\hline

57669.146 & P60r & 144.54 & 26.626 & 25.0 & ab & 0 \\
57670.098 & P48R & 194.09 & 25.027 & 25.0 & ab & 0 \\
57670.124 & J-SPM & 482.61 & 33.45 & 25.0 & vega & 0 \\
57670.124 & Y-SPM & 277.20 & 33.98 & 25.0 & vega & 0 \\
57670.124 & i-SPM & 252.34 & 5.51 & 25.0 & ab & 0 \\
57670.124 & r-SPM & 135.89 & 5.63 & 25.0 & ab & 0 \\
57670.124 & z-SPM & 252.81 & 11.37 & 25.0 & ab & 0 \\
57670.171 & P60r & 162.93 & 6.0025 & 25.0 & ab & 0 \\
57670.173 & P60i & 251.19 & 11.568 & 25.0 & ab & 0 \\
57670.174 & P60g & 25.351 & 4.2029 & 25.0 & ab & 0 \\
57670.225 & P60r & 145.88 & 12.093 & 25.0 & ab & 0 \\
57670.226 & P60i & 244.34 & 51.761 & 25.0 & ab & 0 \\
57671.157 & Y-SPM & 261.57 & 30.79 & 25.0 & vega & 0 \\
57671.157 & J-SPM & 489.32 & 35.17 & 25.0 & vega & 0 \\
57671.157 & i-SPM & 249.11 & 5.44 & 25.0 & ab & 0 \\
57671.157 & r-SPM & 130.25 & 5.63 & 25.0 & ab & 0 \\
57671.157 & z-SPM & 228.034 & 11.46 & 25.0 & ab & 0 \\
57671.225 & P60i & 226.99 & 367.95 & 25.0 & ab & 0 \\
57671.235 & P60r & 125.89 & 12.755 & 25.0 & ab & 0 \\
57671.237 & P60i & 242.1 & 187.31 & 25.0 & ab & 0 \\
57673.188 & P60r & 120.23 & 135.09 & 25.0 & ab & 0 \\
57673.189 & P60i & 222.84 & 28.735 & 25.0 & ab & 0 \\
57673.211 & P60r & 118.03 & 13.045 & 25.0 & ab & 0 \\
57673.213 & P60i & 255.86 & 141.39 & 25.0 & ab & 0 \\
57674.123 & P60r & 131.83 & 18.212 & 25.0 & ab & 0 \\
57674.125 & P60i & 203.24 & 11.231 & 25.0 & ab & 0 \\
57674.208 & P60r & 96.383 & 12.428 & 25.0 & ab & 0 \\
57674.21 & P60i & 237.68 & 78.809 & 25.0 & ab & 0 \\
57675.155 & r-SPM & 85.90 & 4.91 & 25.0 & ab & 0 \\
57675.155 & J-SPM & 512.39 & 35.07 & 25.0 & vega & 0 \\
57675.155 & Y-SPM & 274.15 & 33.16 & 25.0 & vega & 0 \\
57675.155 & z-SPM & 212.81 & 11.81 & 25.0 & ab & 0 \\
57675.155 & i-SPM & 210.66 & 4.98 & 25.0 & ab & 0 \\
57680.204 & P60r & 61.944 & 89.573 & 25.0 & ab & 0 \\
57681.102 & P48R & 88.716 & 14.708 & 25.0 & ab & 0 \\
57682.096 & P48R & 81.658 & 11.282 & 25.0 & ab & 0 \\
57682.139 & P48R & 94.624 & 12.201 & 25.0 & ab & 0 \\
57682.23 & P60r & 64.863 & 12.546 & 25.0 & ab & 0 \\
57682.231 & P60i & 145.88 & 29.56 & 25.0 & ab & 0 \\
57683.093 & P48R & 85.507 & 11.026 & 25.0 & ab & 0 \\
57683.113 & r-SPM & 50.07 & 4.41 & 25.0 & ab & 0 \\
57683.113 & z-SPM & 204.36 & 12.05 & 25.0 & ab & 0 \\

57683.113 & i-SPM & 155.02 & 4.77 & 25.0 & ab & 0 \\
57683.113 & Y-SPM & 330.82 & 41.09 & 25.0 & vega & 0 \\
57683.113 & J-SPM & 562.34 & 34.62 & 25.0 & vega & 0 \\
57683.138 & P48R & 90.365 & 9.15 & 25.0 & ab & 0 \\
57684.151 & J-SPM & 530.15 & 34.01 & 25.0 & vega & 0 \\
57684.151 & Y-SPM & 299.22 & 35.22 & 25.0 & vega & 0 \\
57684.151 & z-SPM & 207.01 & 11.49 & 25.0 & ab & 0 \\
57684.151 & i-SPM & 153.88 & 4.74 & 25.0 & ab & 0 \\
57684.151 & r-SPM & 37.84 & 4.21 & 25.0 & ab & 0 \\
57685.116 & i-SPM & 153.88 & 4.74 & 25.0 & ab & 0 \\
57685.116 & J-SPM & 575.97 & 34.96 & 25.0 & vega & 0 \\
\hline
\end{tabular}}
\resizebox{.3\textwidth}{!}{\begin{tabular}{ccccHHH}
\hline\hline
MJD & Filter & Flux & $\sigma$ & ZP & zpsys & imageid \\
\multicolumn{1}{c}{(days)} & & \multicolumn{2}{c}{(counts/s)} &&  \\
\hline

57685.116 & r-SPM & 39.12 & 4.54 & 25.0 & ab & 0 \\
57685.116 & Y-SPM & 313.04 & 31.72 & 25.0 & vega & 0 \\
57685.116 & z-SPM & 182.97 & 11.74 & 25.0 & ab & 0 \\
57687.117 & Y-SPM & 320.62 & 37.22 & 25.0 & vega & 0 \\
57687.117 & J-SPM & 521.91 & 34.31 & 25.0 & vega & 0 \\
57687.117 & i-SPM & 131.58 & 4.99 & 25.0 & ab & 0 \\
57687.117 & z-SPM & 196.06 & 11.56 & 25.0 & ab & 0 \\
57687.117 & r-SPM & 35.35 & 4.22 & 25.0 & ab & 0 \\
57687.13 & P60r & 39.811 & 31.534 & 25.0 & ab & 0 \\
57687.132 & P60i & 129.42 & 10.728 & 25.0 & ab & 0 \\
57688.112 & r-SPM & 35.74 & 4.32 & 25.0 & ab & 0 \\
57688.112 & Y-SPM & 312.75 & 44.34 & 25.0 & vega & 0 \\
57688.112 & z-SPM & 185.35 & 12.53 & 25.0 & ab & 0 \\
57688.112 & i-SPM & 133.17 & 4.81 & 25.0 & ab & 0 \\
57688.113 & P60r & 40.551 & 34.361 & 25.0 & ab & 0 \\
57688.114 & J-SPM & 548.02 & 35.16 & 25.0 & vega & 0 \\
57688.115 & P60i & 120.23 & 25.469 & 25.0 & ab & 0 \\
57690.114 & J-SPM & 520.47 & 34.29 & 25.0 & vega & 0 \\
57690.114 & z-SPM & 155.88 & 11.60 & 25.0 & ab & 0 \\
57690.114 & r-SPM & 26.91 & 4.03 & 25.0 & ab & 0 \\
57690.114 & i-SPM & 132.80 & 4.80 & 25.0 & ab & 0 \\
57690.114 & Y-SPM & 367.79 & 48.63 & 25.0 & vega & 0 \\
57691.112 & z-SPM & 168.73 & 11.70 & 25.0 & ab & 0 \\
57691.112 & i-SPM & 121.89 & 4.62 & 25.0 & ab & 0 \\
57691.112 & r-SPM & 22.72 & 3.93 & 25.0 & ab & 0 \\
57691.112 & J-SPM & 546.01 & 34.56 & 25.0 & vega & 0 \\
57691.112 & Y-SPM & 301.44 & 34.02 & 25.0 & vega & 0 \\
57692.14 & J-SPM & 487.08 & 33.76 & 25.0 & vega & 0 \\
57692.14 & r-SPM & 30.28 & 4.14 & 25.0 & ab & 0 \\
57692.14 & z-SPM & 145.88 & 11.36 & 25.0 & ab & 0 \\
57692.14 & i-SPM & 123.25 & 4.67 & 25.0 & ab & 0 \\
57692.14 & Y-SPM & 330.52 & 42.12 & 25.0 & vega & 0 \\
57693.15 & J-SPM & 537.53 & 41.39 & 25.0 & vega & 0 \\
57693.15 & r-SPM & 23.86 & 4.80 & 25.0 & ab & 0 \\
57693.15 & z-SPM & 117.49 & 12.87 & 25.0 & ab & 0 \\
57693.15 & i-SPM & 125.77 & 5.44 & 25.0 & ab & 0 \\
57695.108 & P60i & 84.723 & 138.12 & 25.0 & ab & 0 \\
57695.18 & P60i & 91.201 & 21.84 & 25.0 & ab & 0 \\
57696.08 & P60i & 103.75 & 18.156 & 25.0 & ab & 0 \\
57696.082 & P60i & 148.59 & 15.055 & 25.0 & ab & 0 \\
57697.105 & P60r & 30.2 & 54.239 & 25.0 & ab & 0 \\
57697.107 & P60i & 66.069 & 12.779 & 25.0 & ab & 0 \\
57697.109 & P60i & 85.507 & 6.3004 & 25.0 & ab & 0 \\
57700.096 & P60i & 114.82 & 106.81 & 25.0 & ab & 0 \\
57702.148 & P60i & 92.045 & 22.042 & 25.0 & ab & 0 \\
57702.15 & P60i & 83.176 & 16.088 & 25.0 & ab & 0 \\
57705.163 & P60i & 80.168 & 16.244 & 25.0 & ab & 0 \\
57723.104 & r-SPM & 9.72 & 0.77 & 25.0 & ab & 0 \\
57723.104 & Y-SPM & 132.07 & 25.99 & 25.0 & vega & 0 \\
57723.104 & i-SPM & 48.26 & 4.49 & 25.0 & ab & 0\\
57723.104 & z-SPM & 67.48 & 11.35 & 25.0 & ab & 0\\
57723.104 & Y-SPM & 113.76 & 31.42 & 25.0 & vega & 0\\
\hline
\end{tabular}}

\end{table}
\end{landscape}
\iffalse
\begin{landscape}
\begin{table}
	\centering
	\caption{%
		Derived photometry for the four lensed \sn images.  The first error quoted for each measurement is the statistical
		uncertainty that is expected to uncorrelated between epochs.  This includes the expected PSF variations discussed
		in \S\ref{sec:wfcpsf}.  The second error, is the systematic error from the background model fit discussed in the text.
		This is will be correlated for measurements obtained with the same filter.
	\label{tb:resolvflux}}
	\input{resolved_photometry.tex}
\end{table}
\end{landscape}
\fi

%%%%%%%%%%%%%%%%%%%%%%%%%%%%%%%%%%%%%%%%%%%%%%%%%%
\section{Lensing galaxy dust properties and image magnifications}
\label{app:indiv_rv}
In this section, we detail the impact of varying the dust properties for the individual image lines of sight in the lensing galaxy. We analyse the impact that the assumptions have on the inferred magnification. 

\begin{figure*}
\centering
\includegraphics[width=.48\textwidth]{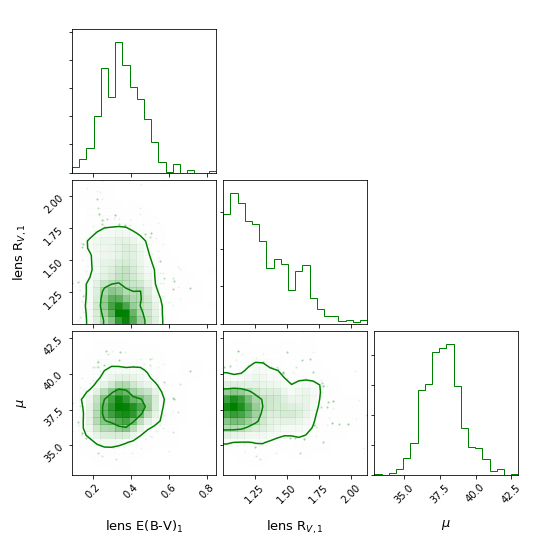}
\includegraphics[width=.48\textwidth]{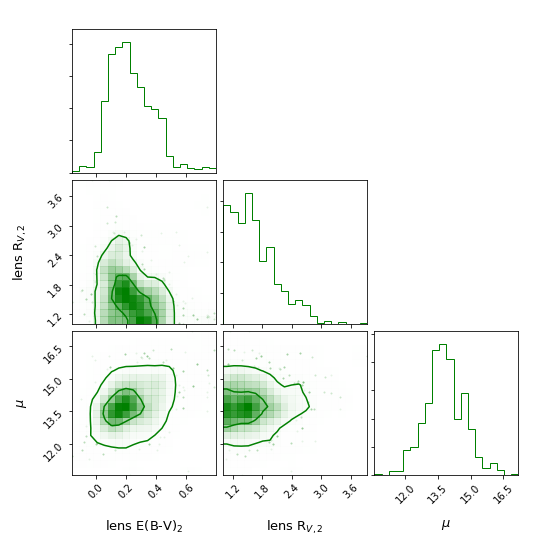}
\includegraphics[width=.48\textwidth]{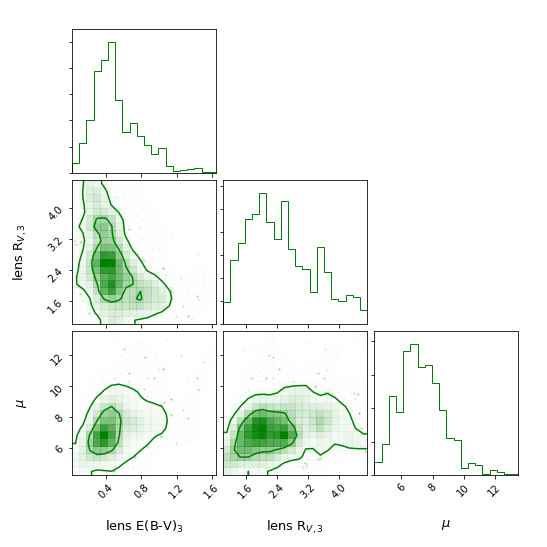}
\includegraphics[width=.48\textwidth]{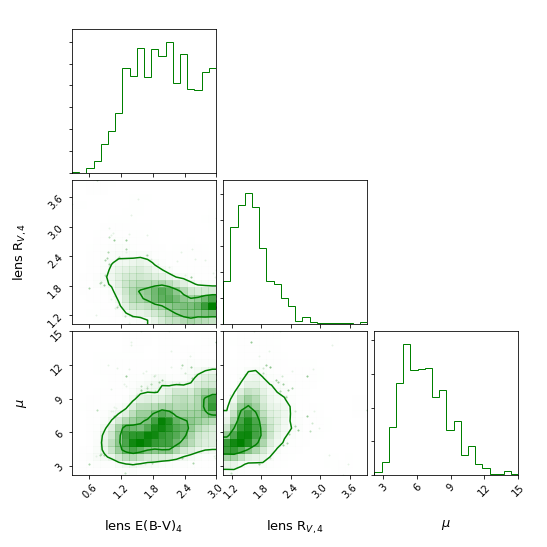}
\caption{Correlation plots between the inferred colour excess $E(B-V)$, total-to-selective absorption ratio, $R_V$, and the magnification, $\mu$ for the images with the $R_V$ as a free parameter for each image.}
\label{fig:indiv_rv_corner}
\end{figure*}
We fit the multiple image model with $R_V$ as a free parameter for each of the images, keeping the host galaxy $R_V$ fixed to the fiducial case of 2. In Figure~\ref{fig:indiv_rv_corner}, we present the correlation between the lens $E(B-V)$, lens $R_V$ and the magnification for each image. While the constraints on the $R_V$ for Images 1, 2 and 3 are not very stringent, we find that there is a very weak correlation between the $R_V$ and $\mu$ for all the images. The inferred total magnification is consistent with the fiducial case of $R_V = 2$.

In addition, we find that the allowed values of $R_V$ for the images do not permit differential extinction between Image 1 and the other Images to explain the discrepancy between the observed and model predicted flux ratios, further suggesting the need for additional lensing from substructures, as discussed in section~\ref{sec:substructure}.

% Don't change these lines
\bsp	% typesetting comment
\label{lastpage}
\end{document}